\documentclass{aastex631}

\usepackage[utf8]{inputenc}
\usepackage{amsmath}
\usepackage[english,bidi=default]{babel}
\babelprovide[import]{arabic}

\usepackage{CJK}

\newcommand\nh{{\em New Horizons}}

\received{}
\revised{}
\accepted{July 24, 2024}

\shorttitle{NH TNO Search}
\shortauthors{Fraser et al.}

\begin{document}

\title{Candidate Distant Trans-Neptunian Objects Detected by the New Horizons Subaru TNO Survey}

\author[0000-0001-6680-6558]{Wesley C. Fraser}
\affiliation{National Research Council of Canada, Herzberg Astronomy and Astrophysics Research Centre, 5071 W. Saanich Rd. Victoria, BC, V9E 2E7, Canada}
\affiliation{Department of Physics and Astronomy, University of Victoria, Elliott Building, 3800 Finnerty Road, Victoria, BC V8P 5C2, Canada}

\author[0000-0003-0333-6055]{Simon B. Porter}
\affiliation{Southwest Research Institute, 1301 Walnut St., Suite 400, Boulder, CO 80302, USA}

\author[0000-0002-9179-8323]{Lowell Peltier}
\affiliation{National Research Council of Canada, Herzberg Astronomy and Astrophysics Research Centre, 5071 W. Saanich Rd. Victoria, BC, V9E 2E7, Canada}
\affiliation{Department of Physics and Astronomy, University of Victoria, Elliott Building, 3800 Finnerty Road, Victoria, BC V8P 5C2, Canada}
\affiliation{Department of Physics \& Astronomy, University of British Columbia, 6224 Agricultural Road, Vancouver, BC V6T 1Z1, Canada}

\author[0000-0001-7032-5255]{JJ Kavelaars}
\affiliation{National Research Council of Canada, Herzberg Astronomy and Astrophysics Research Centre, 5071 W. Saanich Rd. Victoria, BC, V9E 2E7, Canada}
\affiliation{Department of Physics and Astronomy, University of Victoria, Elliott Building, 3800 Finnerty Road, Victoria, BC V8P 5C2, Canada}

\author[0000-0002-3323-9304]{Anne J. Verbiscer}
\affiliation{Department of Astronomy, University of Virginia, P.O. Box 400325, Charlottesville, VA 22904-4325, USA}
\affiliation{Southwest Research Institute, 1301 Walnut St., Suite 400, Boulder, CO 80302, USA}

\author[0000-0003-0854-745X]{Marc W. Buie}
\affiliation{Southwest Research Institute, 1301 Walnut St., Suite 400, Boulder, CO 80302, USA}

\author[0000-0001-5018-7537]{S. Alan Stern}
\affiliation{Southwest Research Institute, 1301 Walnut St., Suite 400, Boulder, CO 80302, USA}

\author{John R. Spencer}
\affiliation{Southwest Research Institute, 1301 Walnut St., Suite 400, Boulder, CO 80302, USA}

\author[0000-0001-8821-5927]{Susan D. Benecchi}
\affiliation{Planetary Science Institute, 1700 East Fort Lowell, Suite 106, Tucson, AZ 85719, USA}

\author[0000-0003-4143-4246]{Tsuyoshi Terai}
\affiliation{Subaru Telescope, National Astronomical Observatory of Japan 650 North A\`{o}hoku Place, Hilo, HI 96720, USA}

\author[0000-0002-0549-9002]{Takashi Ito}
\affiliation{Center for Computational Astrophysics, National Astronomical Observatory of Japan, Osawa 2-21-1, Mitaka, Tokyo, 181-8588, Japan}

\author[0000-0002-3286-911X]{Fumi Yoshida}
\affiliation{University of Occupational and Environmental Health, 1-1 Iseigaoka, Yahata, Kitakyushu 807-8555, Japan}
\affiliation{Planetary Exploration Research Center, Chiba Institute of Technology, 2-17-1 Tsudanuma, Narashino, Chiba 275-0016, Japan}

\author[0000-0001-6942-2736]{David W. Gerdes}
\affiliation{Department of Physics, University of Michigan, Ann Arbor, MI 48109, USA}
\affiliation{Department of Astronomy, University of Michigan, Ann Arbor, MI 48109, USA}

\author[0000-0003-4827-5049]{Kevin J. Napier}
\affiliation{Department of Physics, University of Michigan, Ann Arbor, MI 48109, USA}

\author[0000-0001-7737-6784]{Hsing Wen Lin}
\affiliation{Department of Physics, University of Michigan, Ann Arbor, MI 48109, USA}

\author[0000-0001-8221-8406]{Stephen D. J. Gwyn}
\affiliation{National Research Council of Canada, Herzberg Astronomy and Astrophysics Research Centre, 5071 W. Saanich Rd. Victoria, BC, V9E 2E7, Canada}

\author[0000-0002-7895-4344]{Hayden Smotherman}
\affiliation{DIRAC Institute, Department of Astronomy, University of Washington, 3910 15th Avenue NE, Seattle, WA 98195, USA}

\author[0000-0003-2239-7988]{Sebastien Fabbro}
\affiliation{National Research Council of Canada, Herzberg Astronomy and Astrophysics Research Centre, 5071 W. Saanich Rd. Victoria, BC, V9E 2E7, Canada}
\affiliation{Department of Physics and Astronomy, University of Victoria, Elliott Building, 3800 Finnerty Road, Victoria, BC V8P 5C2, Canada}

\author[0000-0003-3045-8445]{Kelsi N. Singer}
\affiliation{Southwest Research Institute, 1301 Walnut St., Suite 400, Boulder, CO 80302, USA}

\author[0000-0003-0193-9151]{Amanda M. Alexander}
\affiliation{Southwest Research Institute, 1301 Walnut St., Suite 400, Boulder, CO 80302, USA}

\author[0000-0003-1260-9502]{Ko Arimatsu}
\affiliation{National Astronomical Observatory of Japan, Tokyo, Japan}

\author[0000-0002-8236-7396]{Maria E. Banks}
\affiliation{NASA Goddard Space Flight Center, 8800 Greenbelt Rd, Greenbelt, MD 20771, USA}

\author[0000-0002-7277-3980]{Veronica J. Bray}
\affiliation{Lunar and Planetary Laboratory, University of Arizona, 1541 University Blvd, Tucson, AZ 85721, USA}

\author[0000-0002-8262-0320]{Mohamed Ramy El-Maarry}
\affiliation{Department of Earth Sciences, and Planetary Science Research Group, Khalifa University, Abu Dhabi, UAE}

\author[ 0000-0002-2530-0427]{Chelsea L. Ferrell}
\affiliation{Independent}

\author{Tetsuharu Fuse}
\affiliation{National Astronomical Observatory of Japan, Tokyo, Japan}

\author[0000-0002-7851-3186]{Florian Glass}
\affiliation{Independent}

\author[0000-0003-0437-3296 ]{Timothy R. Holt}
\affiliation{Centre for Astrophysics, University of Southern Queensland,  West St, Darling Heights QLD 4350, Australia}

\author{Peng Hong}
\affiliation{Planetary Exploration Research Center, Chiba Institute of Technology, 2-17-1 Tsudanuma, Narashino, Chiba 275-0016, Japan}

\author[0000-0001-7918-118X]{Ryo Ishimaru}
\affiliation{Planetary Exploration Research Center, Chiba Institute of Technology, 2-17-1 Tsudanuma, Narashino, Chiba 275-0016, Japan}

\author[0000-0001-6255-8526]{Perianne E. Johnson}
\affiliation{Institute for Geophysics, Jackson School of Geosciences, University of Texas at Austin,10100 Burnet Road (R2200), Austin, Texas, USA}

\author[0000-0003-3234-7247]{Tod R. Lauer}
\affiliation{NOIRLab, P.O. Box 26732, Tucson, Arizona USA, 85726}

\author[0000-0002-6477-1360]{Rodrigo Leiva}
\affiliation{Instituto de Astrof\'isica de Andaluc\'ia, Consejo Superior de Investigaciones Cient\'{i}ficas, Glorieta de la Astronom\'ia s/n, 18008 Granada, Spain}

\author[0000-0003-0926-2448]{Patryk S. Lykawka}
\affiliation{Kindai University, Shinkamikosaka 228-3, Higashiosaka, Osaka, 577-0813, Japan}

\author[0000-0002-0362-0403]{Raphael Marschall}
\affiliation{CNRS, Observatoire de la C\~ote d'Azur, Laboratoire J.-L. Lagrange, CS 34229, 06304 Nice Cedex 4, France}

\author[0000-0003-0930-6674]{Jorge I. N\'{u}\~{n}ez}
\affiliation{Johns Hopkins University Applied Physics Laboratory 11100 Johns Hopkins Road, Laurel, MD 20723-6099, USA}

\author[0000-0002-9365-7989]{Marc Postman}
\affiliation{Space Telescope Science Institute, 3700 San Martin Drive, Baltimore, MD 21218, USA}

\author[0000-0003-2768-0694]{Eric Quirico}
\affiliation{Universit\'e Grenoble Alpes, CNRS, Institut de Planetologie et Astrophysique de Grenoble (IPAG), UMR 5274, Grenoble, F-38041, France}

\author[0000-0003-2805-4994]{Alyssa R. Rhoden}
\affiliation{Southwest Research Institute, 1301 Walnut St., Suite 400, Boulder, CO 80302, USA}

\author[0000-0001-8994-032X]{Anna M. Simpson}
\affiliation{Department of Physics and Kavli Institute for Astrophysics and Space Research, Massachusetts Institute of Technology, Cambridge, MA 02139, USA}
\affiliation{Department of Physics University of Michigan Ann Arbor, MI 48109, USA}

\author[0000-0003-1316-5667]{Paul Schenk}
\affiliation{Lunar and Planetary Institute (USRA), Houston, TX 77058, USA}

\author[0000-0001-8671-5901]{Michael F. Skrutskie}
\affiliation{Department of Astronomy, University of Virginia, P.O. Box 400325, Charlottesville, VA 22904-4325, USA}

\author[0000-0002-5358-392X]{Andrew J. Steffl}
\affiliation{Southwest Research Institute, 1301 Walnut St., Suite 400, Boulder, CO 80302, USA}

\author[0000-0002-1988-223X]{Henry Throop}
\affiliation{Independent}

\correspondingauthor{Wesley C. Fraser}
\email{wesley.fraser@nrc-cnrc.gc.ca}

\begin{abstract}
We report the detection of 239 trans-Neptunian Objects discovered through the on-going \nh\ survey for distant minor bodies being performed with the Hyper Suprime-Cam mosaic imager on the Subaru Telescope. These objects were discovered in images acquired with either the r2 or the recently commissioned EB-gri filter using shift and stack routines. Due to the extremely high stellar density of the search region down stream of the spacecraft, new machine learning techniques had to be developed to manage the extremely high false positive rate of bogus candidates produced from the shift and stack routines. We report  discoveries as faint as r2$\sim26.5$.  We highlight an overabundance of objects found at heliocentric distances $R\gtrsim70$~au compared to expectations from modelling of the known outer Solar System. If confirmed, these objects betray the presence of a heretofore unrecognized abundance of distant objects that can help explain a number of other observations that otherwise remain at odds with the known Kuiper Belt, including detections of serendipitous stellar occultations, and recent results from the Student Dust Counter on-board the \nh\ spacecraft. 
\end{abstract}


\section{Introduction \label{sec:intro}}

NASA's \nh\ spacecraft launched in January 2006 and carried out a successful fly-through of the Pluto system in July 2015 \citep{Stern2015} and continues to be fully operational. 
In 2004, two years before launch, the \nh\ team began a ground-based observational search to identify a Kuiper Belt Object (KBO) that the spacecraft could fly by at close range after the Pluto encounter.  Due to Pluto's location, the trajectory of \nh\ sent it toward the apparent position of Sagittarius, near the galactic center. Any possible encounter target would be located along or near the trajectory of \nh\ beyond Pluto. In 2009, any potentially encounterable KBOs had sky locations that projected to 0 degrees Galactic latitude and 12 degrees longitude.  The dense stellar-background created by the alignment of the projected sky location of the spacecraft trajectory with the galactic centre makes searches very difficult owing to contamination of the KBOs as they inevitably traverse background stars.  Likewise, most trans-Neptunian Objects (TNOs) are small (most have diameters $<200$~km), distant ($>30$~au) and faint ($m_{R}<25.0$~mag R-band), and therefore require large telescopes to discover them. Searches therefore used the largest telescopes available paired with instruments having large fields of view. 
    
The high stellar background densities thwarted the initial observation campaigns conducted in 2004-2005 from Subaru using its Suprime-Cam imager \citep{Miyazaki2002}. 
By 2011 the target field had moved to a slightly less dense background, and the apparent locations of TNOs downstream of the spacecraft occupied a smaller area on the sky \citep{Buie2023}. The team undertook a new search, employing a combination of facilities, including Subaru and its Suprime-Cam imager \citep{Miyazaki2002}, the Canada-France-Hawaii Telescope MegaPrime imager \citep{Boulade2003}, and Megacam \citep{McLeod2015} and IMACS \citep{Dressler2006} on the Magellan telescopes. 
In 2014, the Hubble Space Telescope (HST) conducted a highly targeted search, resulting in the discovery of (486958) Arrokoth (2014 MU69), and enabling the first flyby of a cold classical Kuiper Belt Object in January 2019 \citep{Stern2019}. The results of these early searches are reported in \citet{Buie2023}.

After the successful Arrokoth flyby, \nh\ remains healthy and continues on a trajectory through the trans-Neptunian region, maintaining the capability for the spacecraft to observe additional TNOs, with current heliocentric distance of $R\sim59$~au. 
Beginning in 2018, a renewed search began in earnest to identify targets, not only for close encounter flyby but also for observation by the spacecraft at distances, resolutions and phase angles not possible from Earth-based observatories. The first survey was with the Dark Energy Camera on the Blanco Telescope \citet[program 2018A-0177, PI A. Parker][]{Flaugher2015}. 
The 8.2-m Subaru Telescope with its 1.77 deg$^{2}$ Hyper Suprime-Cam (HSC) \citep[Subaru-HSC;][]{Miyazaki2018} field of view is an excellent facility for carrying out such a search and so, in 2020 
we began using that camera. At first we observed with the facility r2 filter, and in fall 2022, commissioned an extra wide ``EB-gri" facility filter, which allows the search to detect objects $\sim$0.5 magnitudes fainter than with the r2 filter. Depending on the night, our survey reached $26^{th}$-$27^{th}$~magnitude, even at the high stellar densities of the near-galactic plane search fields. 
    
The following sections describe the survey strategy, the Subaru-HSC instrument as applied to this program,  the data reductions,  and the novel search techniques developed using machine learning tools to optimize object discovery. We finish by discussing the detections found beyond the nominal ``edge'' of the Kuiper Belt ($R\sim50$~au) thus far and the implications they have for the \nh\ mission as well as for Kuiper Belt science in general. 
 
\section{The Camera and Observing Strategy  \label{sec:obs_strategy}}
In this section, we describe the observing program, including a description of the camera we heavily utilized for the search (\ref{sec:HSC}), and the observing strategy used to provide both discovery and tracking observations (\ref{sec:sub_obs_strategy}). 

\subsection{Subaru-HSC and the EB-gri filter\label{sec:HSC}}

The Subaru-HSC is a wide-field camera whose focal plane is populated with a mosaic of 116 CCDs, 103 of which are available for science.
The camera has a pixel size of 15 $\mu$m and a pixel scale of $0.168\arcsec$. The $\sim 1.5^\circ$ field of view camera is designed to conduct large-scale surveys of the sky with significant depth and area to high photometric and astrometric precision.
The delivered image quality is excellent across the field of view, with a median seeing in the $i$-band of about $0.6''$ \citep{aihara2019,aihara2022}.
For technical details of HSC, see
\citet{Miyazaki2018} for system design,
\citet{komiyama2018} for camera dewar design,
\citet{kawanomoto2018} for filter system,
and \citet{furusawa2018} for  on-site quality-assurance system.\footnote{
the dedicated Subaru-HSC web-page is available at \url{https://www.subarutelescope.org/Observing/Instruments/HSC/}.}

\begin{figure}
    \includegraphics[width=0.95\textwidth]{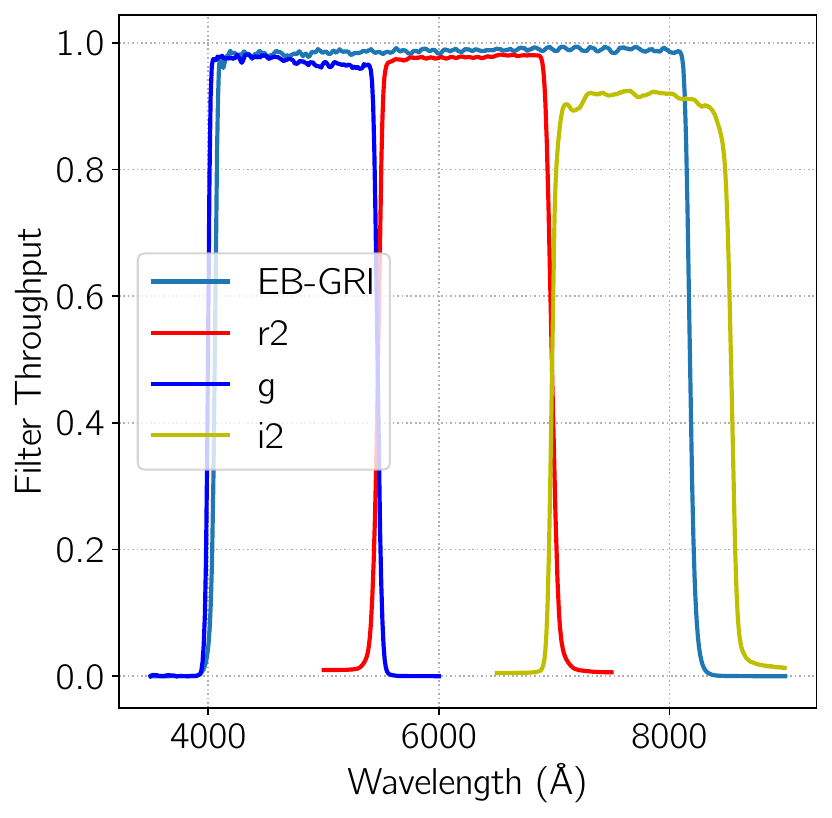}
    \caption{ The bandpasses of the g, r2, i2, and EB-gri filters currently in use with the HSC. \label{fig:filters}}
\end{figure}

As part of this project, our group commissioned a new broad-band EB-gri filter that roughly spans the g, r2, and i2 filters provided by Subaru (see Figure~\ref{fig:filters}). This is similar to the ``w-band'' (GRI.MP9605) filter in use on CFHT-Megacam.\footnote{\url{https://www.cfht.hawaii.edu/Instruments/Imaging/MegaPrime/specsinformation.html}} From photon noise considerations, we anticipated a $\sim0.5$~mag increase in equivalent r survey depth compared to use of the r2 filter. We provide estimates of r- and g-band color terms in the appendix.

\subsection{The Observing Strategy \label{sec:sub_obs_strategy}}

\begin{figure}
\includegraphics[width=0.95\textwidth]{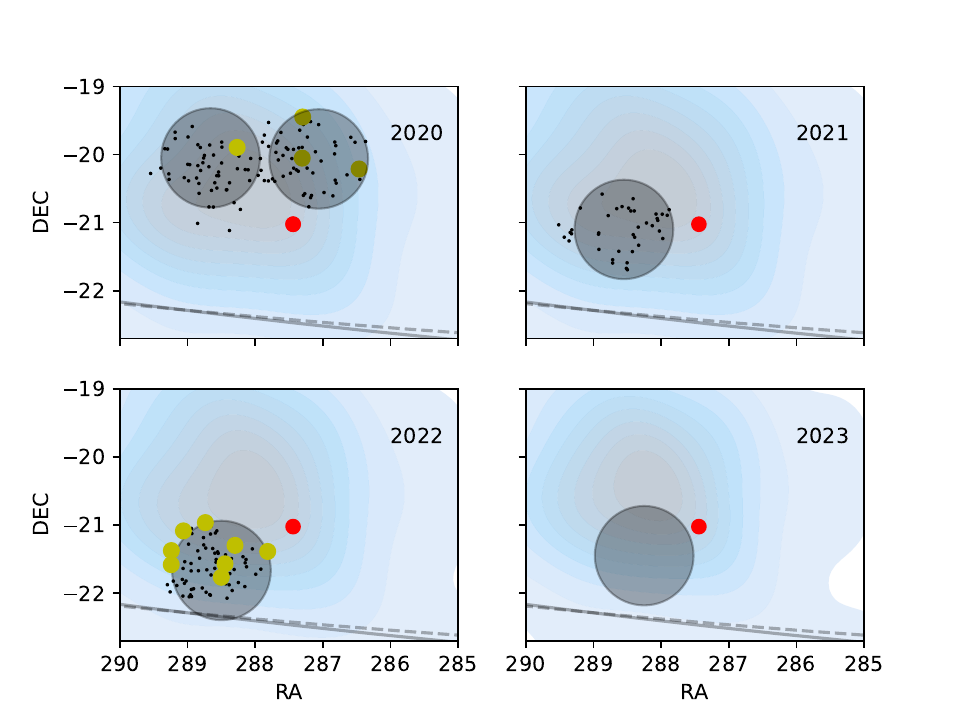}
    \caption{Density of NH-observable TNOs (contours and blue/grey background) detectable with Subaru/HSC expected from the OSSOS++ model.
    Grey circles show HSC search fields for each year 2020-2023A. 
    The solid line indicates the location of the ecliptic plane while the dashed line indicates the invariable plane at 45~au.
    The red dot marks the location of the bright star $\pi-Sgr$ which creates a zone of avoidance. 
    Yellow circles show locations of Subaru-discovered TNOs that were either subsequently observed by (in 2020-2021), or potentially observable by, the NH-LORRI instrument. In each panel, the black points show the discoveries made with observations of that year.  The 2023A HSC deep search field, centered at (RA,DEC) = (289.1, $-20.36$), went $\sim0.5$ mag deeper than searches before 2022B, and some TNOs that were near bright stars then had moved into clearer sky in 2023A. This 2023A data has been used for linkages to earlier detections, and the search for new targets is ongoing. \label{fig:obslayout}}
\end{figure}

The primary objective of these observing campaigns is to discover and track TNOs whose space locations will be accessible by \nh\, enabling a close fly-by.   
TNOs that were too far away to allow an encounter could be observed at larger distances, some at resolutions that exceed what might be achievable from Earth-localized observers. Those observations are useful for distant high phase angle and satellite searches.
We utilize TNO orbital and size distribution models to determine the optimal sky search location based on computed encounter trajectories between model objects and the spacecraft trajectory. 
The current sky locations of model objects with which \nh\ could have a close encounter, allowing for small course corrections ($\Delta-V \sim 100s$~m/s), determined the optimal search area.
The TNOs discovered in these searches were not encounterable, but many were close enough to be observable using the \nh\ LORRI camera \citep{2008SSRv..140..189C}.

The orbit model substantially influences the outcome of these searches. 
For planning the initial searches for Arrokoth \citep[2011-2012,][]{Porter2022}, the bulk orbital properties of the known classical KBOs were used; these searches did not result in the discovery of an encounter target but did discover numerous TNOs that could be observed remotely.
The HST-based search in 2014 \citep{Buie2023} transitioned to using debiased orbit distribution models from the Canada-France Ecliptic Plane Survey \citep[CFEPS  L7-model;][]{Petit2011}. 
The discovery of Arrokoth, a few arc minutes from the predicted optimal search location, confirmed the veracity of the model-based search field.

In 2020, after \nh\ successfully encountered Arrokoth in 2019 \citep{Stern2019}, we began a new campaign using the Subaru-HSC to search for new targets observable with the \nh\ LORRI.
The exact telescope pointings were determined using a debiased population model derived from the orbital and size distributions determined by the Outer Solar System Origins Survey \citep[OSSOS,][]{Bannister2018,Kavelaars2021}.
The transition to the OSSOS++ model was triggered by the evolving understanding of the shape of the size-distribution of TNOs, which were strongly informed by the cratering records on Pluto and Arrokoth \citep{Singer2019, Spencer2020} and the OSSOS measured population distributions in distant Neptune mean-motion resonances \citep{Crompvoets2022} and includes the latest measurements of the absolute magnitude distributions from OSSOS \citep{Kavelaars2021, Petit2023}. 
The wide latitude and faint magnitude distributions of potential targets, caused by the broader inclination distribution of these distant objects, necessitated a wide field search for faint TNOs, making Subaru-HSC an excellent search platform for the task. 
With the sample of observable model objects determined, we computed ephemerides and determined the optimal sky locations for each night of allocated observing. 
These optimal sky locations shifted each night to ensure maximal recovery of the sample of observable TNOs, as predicted by our model inputs. See Figure~\ref{fig:obslayout} for examples of the observing layouts used in 2020 through 2023. Each night followed essentially the same observing strategy: acquiring as many consecutive 90~s exposures as possible within the allotted time. The sequence was broken only for refocusing exposures or loss of tracking.
The post-Arrokoth searches provided a sample of 239 TNOs (submitted to the Minor Planet Center) whose observed orbital distributions can be used to test the completeness of the model used in planning those surveys. 

\section{Reductions and Search Techniques\label{sec:search}}

This section describes the pipeline steps required to go from raw pixels to detected TNOs. 
These steps include image preprocessing and calibrations (\ref{sec:preproc}), the shift and stack detection techniques (\ref{sec:sns}), culling of bad candidate sources using machine learning techniques and human confirmation of good sources (\ref{sec:ml}), and finally intra-night source linking to verify detections (\ref{sec:linking}). 

\subsection{Image Preprocessing  \label{sec:preproc}}

Raw imagery was preprocessed and calibrated using the LSST pipeline v19\footnote{As the results were satisfactory for our needs and to maintain consistency, we chose to stick with LSST pipeline v19 throughout the project rather than update to newer releases from year to year.} \citep{Bosch2018,Bosch2019}. For each night, an average bias frame was constructed from bias frames acquired during the twilight of that night, using the \verb|constructBias| task. An average flat field for the night was generated using the archival dome flats that were temporally closest to the night in question (usually acquired in the previous dark run), using the 
\verb|constructFlat| task. Using the average flats and biases constructed for each night, the \verb|processCCD| package was used to debias, flat-field, and perform photometric and astrometric calibrations. Calibrations were performed against the merged PS1-GAIA catalogue,\footnote{\url{https://community.lsst.org/t/pan-starrs-reference-catalog-in-lsst-format/1572}} resulting in $\sim0.02$~mag RMS photometric calibration quality and $\sim0.04\arcsec$ RMS astrometric calibration quality. Similar astrometric quality was found with the EB-gri filter, though with a few tenths of a magnitude worse absolute photometric calibration. Roughly $\sim2\%$ of cases were catastrophically compromised through causes such as guider loss or mid-exposure focus changes, though with the low occurrence rate; this held no significant bearing on our survey depth.

To enable the characterization of the detection efficiency as a function of brightness and orbital parameters, a significant number of artificial moving bodies were implanted in the data. Objects were randomized across a range of semi-major axes, eccentricities, and inclinations, and with randomized absolute magnitudes to result in brightnesses $20\leq m \leq 28$ in the observed filter. Between 20 and 40 sources were generated per chip. Point spread function (PSF) characterization was done with the TRailed Imagery in Python (TRIPPy) package \citep{Fraser2016}. All non-saturated sources were examined for low ellipticity and no neighboring sources within 8~FWHM, and those with $<0.5$~mag difference between their Kron \citep{Kron1980} and circular aperture magnitudes were used to generate the PSFs. The PSF lookup tables were then cleaned of any residual signal by replacing pixels that had variations more than $2\sigma$ away from the mean and outside 2~FWHM radius, with a random value drawn from the distribution of residual-free pixels. This later step was necessitated by the highly crowded fields but resulted in PSFs of sufficient fidelity for our needs. 

Source injection was also done with TRIPPy. Each source was injected with trailing matching its simulated orbital trajectory. Poissonian noise was added.

The LSST pipeline was used to produce difference images of the implanted imagery. The \verb|makeDiscreteSkyMap|, \verb|makeCoaddTempExp|, and \verb|assembleCoadd| routines were used to generate the subtraction template, itself generated from the 30\% of images from a night with the best image quality (IQ). The frames used in the template generation should span the full temporal range of the nightly sequence. This generally resulted in excellent removal of bright stationary sources without catastrophic removal of the flux from moving sources. The subtraction was performed with \verb|imageDifference|. The full template and subtraction routine was done on a nightly basis, with each CCD treated individually. The latter choice was made due to some bugs discovered during the image subtraction process that manifests only when multiple CCDs were considered together during the template generation and subtraction phases. 

For the 2020 and some of the 2021 of the implanted and subtracted imagery, pixels with large subtraction residuals (variance $>25\times$ the variance of source-free regions) were masked in order to reduce the number of spurious detections and reduce the visual inspection load.  Starting part way through 2021, however, this masking step was removed as it was found that the ResNet (see Section~\ref{sec:ml}) we made use of during the vetting step was able to remove the vast majority of bogus sources resulting from these poorly subtracted regions. 

\subsection{Shift and Stack  \label{sec:sns}}
Subtracted data were searched for moving point sources using a shift-and-stack technique, whereby the nightly frame sequences are shifted at linear rates of motion to counteract the movement of the bodies of interest and then digitally stacked. Near a motion vector that matches the apparent movement of a source, said source appears as a near-point source in the stack, allowing simple detection and achieving depths comparable to the limit of the sidereal stack. 

The first search effort of data observed in 2020 consisted of classical CPU stacking methods \citep[e.g.,][]{Fraser2008, Fuentes2009}, and vetting of many thousands of candidate sources with a large team of volunteer vetters.\footnote{for whom we are forever grateful} This enabled a quick search but with poor performance and extreme human and computational cost. It was quickly recognized that such a method was untenable, and so we turned to more modern stacking and vetting techniques. The techniques implemented evolved from run to run to adjust for lessons learned as we went.

The Kernel-Based Moving Object Detection \citep[KBMOD;][]{Whidden2019, Smotherman2021} software package was used to perform the shift-and-stack search for the 2021 and later data. This software uses graphics processing units to perform the computational heavy lifting, reducing the stack time from days per chip (even on a supercomputer) to hours. For near opposition fields, a grid of 35 rates uniformly spanning $0.5<v<4.0~\mbox\arcsec$/hr -- corresponding to heliocentric distances $30\lesssim R \lesssim 300$~au -- and 15 angles uniformly spanning $-20 \leq a \leq 20^\circ$ from the ecliptic to be sensitive to objects on prograde orbits. The grid spacing was chosen to ensure that an object's flux was spread over a region no wider than $1.4\arcsec$, or roughly twice the typical seeing over Maunakea. It is possible that the use of a finer grid would result in a fainter search depth, but comes at the cost of increased false positive rate and compute time, and so this spacing was chosen as a practical balance between these considerations.

The KBMOD likelihood threshold was set to 5, an ambitious level at which the majority of candidates are noise sources. This threshold resulted in 3,000-5,000 candidate sources per CCD.

For each candidate source, mean and median stacks were produced at the rate and angle that maximized the likelihood as reported by KBMOD. The stacking routine ignored all pixels flagged by the LSST pipeline as \texttt{EDGE}, \texttt{NO\_DATA}, \texttt{SAT}, \texttt{INTRP}, \texttt{REJECT}, which are pixels marked as on an edge, containing no real data, saturated, interpolated over, or rejected by the pipeline for other reasons. These stacks acted as the main products for training and execution of the machine learning tools, which we discuss in the next section.

For the October 2022 data, which were observed a few days before or after the stationary point at $R\sim45$~au, the grid of rates and angles was adjusted to $0.5<v<3.0\arcsec$/hr and $-60 \leq a \leq 60^\circ$ to be sensitive to the objects of interest. The sign of the rates grid was adjusted accordingly for observations that occurred after the stationary point. 

\subsection{Vetting Through Machine Learning\label{sec:ml}}

After the initial quick search of the 2020 data using a team of humans, we implemented machine-learning techniques to provide an initial cull of bad sources from the candidate list, thereby saving enormous amounts of human cost in the final vetting.

The process of vetting is essentially one of binary classification, where a source is labeled as good and kept, or bad and rejected. As such, vetting of shift-stacks represents a task ideally suited for convolutional neural networks. As the search commenced, many different neural networks were explored. Our experimenting included basic convolutional neural networks (CNNs) of varying complexity and depth. 
The 2020 data (nights 03071, 03072, 03093, 03147, 03148; see Table~\ref{tab:deteff}) were searched using
basic CNNs. An ensemble of three identical CNNs that was trained (we discuss the training routine below) using all implanted artificial sources was first used, but with sub-par performance of the network within $\sim1$~magnitude of the nominal detection limit of $r_2\sim27$~mag. As such, a second ``deep'' ensemble of 3 CNN networks was trained using only implants with $r_2>25.5$. This deep network resulted in significantly increased performance at the faint end of the implants. 

We further explored other networks, including ensembles of the above two networks, and networks trained with implanted sources divided in 1.5~mag-wide bins. Only minimal performance gains were found, however, not warranting a re-search of the data. 

\begin{figure}[!ht]
    \includegraphics[width=0.95\textwidth]{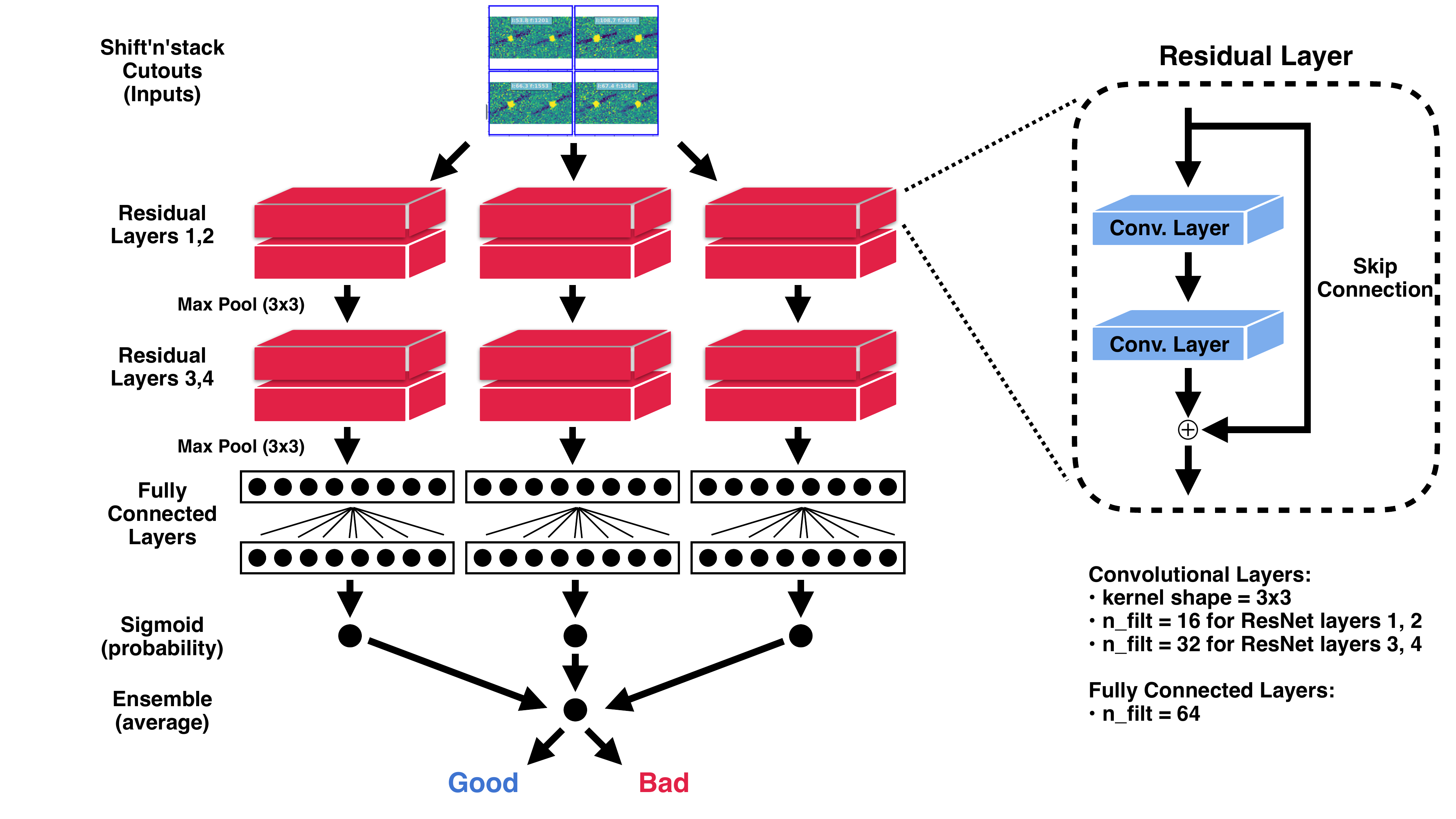}
    \caption{A diagram of the ResNet structure we have adopted. A \emph{master} network consists of ensembles of the deceptively simple network. Each \emph{branch} of the ensemble consists of two residual layers, with each layer consisting of two standard convolutional layers (3x3 kernels, 16 and 32 filters for first pair and second pair of convolutional layers, respectively) with surrounding skip connection and max-pooling between residual layers. The two residual layers are followed by two fully connected layers with 64 nodes each, and finally a sigmoid output. The outputs of the three branches, each with identical structure, but independent training runs, are averaged for the classification probability of this ensemble. The final probability we adopted is the ensemble of three separate \emph{master} ensemble networks like that shown above, but with each ensemble trained using an independent training dataset. We direct the reader to \citet{He2016} for an understanding of the operations of a ResNet. \label{fig:resnet}}
\end{figure}

A remarkable performance boost was found when we implemented Residual Networks \citep[ResNets;][]{He2016}. Our adopted ResNet (see Fig~\ref{fig:resnet}), which was trained on all detected implants regardless of brightness, met or outperformed all previous networks as measured by depth and peak detection efficiency, and came with significantly reduced numbers of trainable parameters compared to all other networks we tried. This is important for overfitting considerations, which we discuss further below.

We adopted an ensemble of networks \citep[e.g.,][and references therein]{Rosen1996} of identical structure and trained three \emph{master} networks independently. Each master network itself is an ensemble of separate singular networks which we call \emph{branches}. Each branch of a master is trained on the same data, but initialized separately. The final ensemble prediction probability is an average of prediction probabilities from each master network, each itself outputting an ensemble of probabilities from its singular networks.

To train each network (all CNNs and ResNets) we utilized supervised learning. Determining a successful training procedure was an extremely difficult task. The successful procedure we found was as follows. A candidate KBMOD source detection was labelled good if it was within 3 pixels of a known implanted source location with rate of motion within $0.2\arcsec$/hr of the implant's rate of motion. All other sources were labelled as bad. The full candidate sample was vastly dominated by approximately 200:1 junk candidates to good detections, which is much too unbalanced for training. As such, a random subset consisting of all good detections, and a subset of the junk detections of size $1.6\times$ the good detections was generated. Any more unbalancing of the sample towards junk detections would result in failure to train. We utilized a sample weighted loss, with sample weight of 0.7 for junk candidates and 1 for good sources, as these values resulted in excellent network outputs. Training proceeded through usual means: the ADAM\footnote{ADAM is not an acronym; it is a name given to the optimization routine.} minimizer was used for 200 epochs \citep{Kingma2017}, with a batch size of 256 (any smaller resulted in failed training), and a categorical entropy loss function given by $\log L = \frac{1}{N} \sum_{i=1}^{N} (1-y_i) \log(1-p_i) - y_i \log(p_i)$ where $p_i$ is the probability that sample $i$ is good, and $y_i$ is the ground truth for sample $i$, equalling 1 if the sample is good, 0 if bad. 

Experiments with this training demonstrated small variations in performance from training run to training run even with the same training set, and so from a given training set, an ensemble of three \emph{branch} networks was trained, but each with a different random initialization resulting in slightly different classification probabilites from each branch. These three commonly trained branch networks make up one \emph{master} network (see Figure~\ref{fig:resnet}). 

To ensure the ability to reject the broad range of junk sources that appear in the data set, multiple unique randomized training sets were generated as described above, with the same set of good sources, but a newly randomized set of junk sources. For each unique training set, a master network was trained. After training $\sim10$ master networks on different randomized training sets, the best performing three were accepted for production. 

The training data consisted of the KBMOD outputs from all chips of a field, from nights 03072 and 03093 (see Table~\ref{tab:deteff}). After generation of the random set, but before training, shift and rotation augmentations were implemented, whereby a given source was duplicated, but with 3 separate cardinal rotations, and as well, shifted by 1 pixel left, right, up, and down producing a training sample of size 419,605 samples. This augmentation resulted in a factor of $\sim2$ reduction in mislabels of junk sources. 

To ensure a satisfactory search depth, it was required to select a unique acceptance threshold for each night separately. This value (usually between 0.05 and 0.6) was manually tuned to select the point where a reduction in threshold did not result in an appreciable increase in depth, but did result in a significant increase in number of sources requiring human vetting. For a few nights, the performance of the network was less than satisfactory. These circumstances seemed to result from highly variable seeing conditions, or poor subtractions, making the stacks quantitatively different than those used for the training. One solution would be to train a network with a more diverse range of observational conditions. We reserve this exploration for future work.

The performance of the network ensemble was admirable. The exact performance varied from field to field, but typically the use of the ensemble ResNet resulted in a $\sim 500 - 1000\times$ decrease in bogus sources in the candidate source list, with a loss of $\sim0.2$~mag in total depth compared to the raw KBMOD outputs. Without such a decrease in rate of bogus sources, the final human vetting task would be practically impossible.

The sources labelled as good were passed to a human for final visual vetting. For each source, the mean and median stacks were displayed side-by-side and labelled with the KBMOD likelihood and flux estimates. The junk:good purity at this stage was usually better than 3:1, and a field could be vetted by a trained operator in less than a day. 

The list of human vetted sources was typically of order $\sim 200$ sources long. Above likelihood $\sim7$ all sources labelled good by a human were obviously real. The final human-vetted list contained some bright false positives caused primarily by subtraction residuals of bright stars. The list however, was  dominated with low flux sources with likelihoods $<7$, many of which were not real. These sources were predominately driven by the extremely high stellar density and consequent high density of subtraction residuals. In other words, nightly false positive rates were high because of our near--galactic-plane observing. Fortunately, a final false positive rejection occurred during the linking process, which we describe below. 

One major concern we had with the use of machine learning was overfitting. Overfitting can come in multiple flavours. One particularly damaging form is one in which the network is able to identify injected sources better than real sources, which leads to an incorrect measure of detection efficiency. Detection of overfitting in neural network training is notoriously difficult and is an ongoing field of study. We are confident that overfitting is not an issue from two findings in our search. Namely, real objects are discovered across different nights of the search with a frequency that appears consistent with expectations from the nightly detection efficiencies. As well, we found that the searches of the majority of nights not used in the training perform equally as well as those used during the training by measures such as peak detection efficiency, limiting magnitude, number of real detections, and number of junk detections requiring human rejection. That is to say, with few exceptions, the ResNet seemed to perform equally well across many nights of data, not just those for which it was trained. This implies that no sensitivity to peculiarities of those nights -- which may include processing or PSF oddities, for example -- were not what the machine latched onto, but rather, properties of the HSC data in general. We also point out that during training, the number of observations (unique pixels) in the training set was more than $50\times$ larger than the number of trainable parameters, giving confidence that overfitting was extremely unlikely. It seems that if  overfitting did occur, its influence on the results must be minor. The similarity in performance of the network between nights used for training and those other nights which served as a fully independent validation set supports our conclusion that overfitting is not significant. The final detection efficiencies after vetting shift-stack sources with a neural network, and then human vetting are shown in Table~\ref{tab:deteff}.

We highlight that the above technique is a more advanced version of the neural network implemented to vet KBMOD sources presented by \citet{Smotherman2021}. They utilized the ResNet50 \citep{He2016} to cull sources from some DECam shift-stack data, but used less sophisticated training techniques and Gaussian PSFs for implants. The simpler model we present here outperforms this prior attempt and avoids overfitting risks incurred with the huge increase in trainable parameters in the ResNet50 model.

\startlongtable
\begin{deluxetable*}{llllccccccc}
\tabletypesize{\scriptsize}
\tablecaption{HSC Observing Epochs.}

        \tablehead{
        \colhead{Observation} &
        \colhead{Pipeline} &
        \colhead{R.A.} &
        \colhead{Dec.} &
        \colhead{Number} &
        \colhead{Sequence} &
        \colhead{$A$}  &
        \colhead{$c$} &
        \colhead{$M_o$} &
        \colhead{$\sigma$} &
        \colhead{Network}\\
        \colhead{Time (MJD)} &
        \colhead{Night} &
        \colhead{$(^\circ)$} &
        \colhead{$(^\circ)$} &
        \colhead{Exposures} &
        \colhead{Duration (hrs)} &
        \colhead{}  &
        \colhead{} &
        \colhead{} &
        \colhead{} &
        \colhead{Type}
}
\startdata 
58995.4 & 03068 & 287.37585 & -20.22724 &  81 & 2.7 & - & - & - & - & - \\ 
58997.5 & 03070 & 288.74057 & -20.57457 &  76 & 2.7 & - & - & - & - & - \\ 
58998.5 & 03071 & 287.33743 & -20.12725 & 123 & 4.1 & 0.6757 &  0.0070 &  26.6652 &  0.2205 & c \\ 
58999.5 & 03072 & 288.71466 & -20.37960 & 127 & 4.3 & 0.7570 &  0.0056 &  26.4101 &  0.2486 & c \\ 
59000.5 & 03073 & 287.31192 & -20.12834 & 126 & 4.2 & - & - & - & - & - \\ 
59001.5 & 03074 & 288.68802 & -20.38238 &  58 & 4.1 & - & - & - & - & - \\ 
59019.5 & 03092 & 287.14477 & -20.04175 & 129 & 4.4 & - & - & - & - & - \\ 
59020.5 & 03093 & 288.66147 & -20.05024 & 117 & 4.0 & 0.8295 &  0.0054 &  26.3803 &  0.2016 & c \\ 
59021.5 & 03094 & 287.11257 & -20.05892 & 113 & 3.8 & - & - & - & - & - \\ 
59022.5 & 03095 & 288.62929 & -20.05384 & 101 & 3.4 & 0.8568 &  0.0010 &  26.3944 &  0.1838 & f \\ 
59024.5 & 03097 & 287.06207 & -20.06483 & 129 & 4.3 & 0.8192 &  0.0006 &  26.2047 &  0.1815 & f \\ 
59025.5 & 03098 & 288.57787 & -20.06015 & 119 & 4.0 & - & - & - & - & - \\ 
59073.3 & 03145 & 286.25079 & -20.16488 & 116 & 3.9 & - & - & - & - & - \\ 
59074.3 & 03146 & 287.67551 & -20.12296 & 128 & 4.3 & - & - & - & - & - \\ 
59075.3 & 03147 & 286.22292 & -20.16881 & 127 & 4.3 & 0.7642 &  0.0085 &  26.3015 &  0.1758 & f \\ 
59076.3 & 03148 & 287.74292 & -20.16470 & 121 & 4.3 & 0.7841 &  0.0091 &  26.5365 &  0.1989 & f \\ 
59374.5 & 03447 & 288.54961 & -21.09995 & 115 & 4.2 & 0.8214 & 0.0 & 26.6310 & 0.4176 & r \\ 
59382.5 & 03455 & 288.44956 & -21.10000 &  94 & 3.4 & 0.8302 & 0.0 & 26.5853 & 0.2644 & r \\
59400.4 & 03473 & 288.00157 & -20.28053 & 117 & 4.1 & 0.8826 & 0.0039 & 26.9686 & 0.2573 & r \\
59462.2 & 03535 & 288.35953 & -21.35001 & 119 & 4.0 & - & - & - & - & -  \\
59463.2 & 03536 & 286.58291 & -21.34999 & 116 & 3.9 & - & - & - & - & -  \\
59732.5 & 03805 & 288.50000 & -21.66665 & 111 & 3.9 & 0.8343 & 0.0 & 26.6119 & 0.2941 & r \\
59733.5 & 03806 & 288.49996 & -21.66668 & 119 & 4.1 & 0.8275 & 0.0 & 27.0161 & 0.3851 & r \\
59759.5 & 03832 & 288.25939 & -21.42730 & 116 & 3.8 & 0.8902 & 0.0044 & 26.7224 & 0.2565 & r \\
59760.5 & 03833 & 288.25943 & -21.42732 & 114 & 3.9 & 0.8887 & 0.0055 & 26.3220 & 0.3435 & r \\
59872.2 & 03945 & 288.42499 & -21.40842 &  56 & 2.0 & - & - & - & - & - \\
59873.2 & 03946 & 288.42503 & -21.40844 &  65 & 2.2 & 0.8090 & 0.0078 & 26.32 & 0.2532 & r \\
59875.2 & 03948 & 288.37854 & -21.41377 &  64 & 2.1 & - & - & - & - & - \\
60196.3 & 04269 & 288.25001 & -21.44998 & 125 & 4.4 & 0.8208 & 0.0000 & 25.78 & 0.7214& r \\
60197.3 & 04270 & 288.25002 & -21.44998 & 136 & 4.5 & - & - & - & - & - \\ 
\enddata
\tablecomments{
List of observing sequences gathered and the survey discovery efficiency. Nights without reported discovery efficiencies were not searched, but used only for follow-up. Reported detection efficiencies were fit using the functional form $\eta(m) = \frac{A - c(m-20)^2}{1+e^{\frac{m-M_o}{\sigma}}}$ \citep{Bannister2018}.  $A$ is the peak efficiency, at $m=20$. This function was required over simpler representations such as the tanh representation \citep[e.g.][]{Fraser2008} to accurately represent the detection efficiency which has a multimodal behaviour that exhibits a gradual decrease in efficiency with increasing magnitude (quantified by parameter $c$) followed by much more precipitous decline nearer the limiting magnitude (quantified by parameter $M_o$, with a width quantified by $w$).  The network type column describes which type of network CNN (c), flux sorted CNN (f), or ResNet (r) was used for initial KBMOD source rejection. Nights marked with `-' were not searched directly, but used only for follow-up and linking. For all sequences, exposure times were 90~s. Field 03833 encountered an issue whereby the efficiency dropped to $\sim0\%$ brighwtward of $r\sim22$ and so the efficiency parameters quoted apply to only fainter values. These observations were acquired in Subaru programs S20A-OT04, S20B-TE019-G, S20B-081, S20B-TE204-K, S20B-OT04, S21A-OT04, S21A-TE216-K, S21B-OT04, S22A-TE049-K, S22B-TE071-K, and S23B-OT55.
\label{tab:deteff}
}
\end{deluxetable*}

\subsection{Astrometry and Linking \label{sec:linking}}

Turning our initial detections into confirmed TNOs with measurable orbits required a process to expand the single-night detections into multi-night arcs.
This arc-expansion process was necessarily less automatic than the initial detection process, owing both to the idiosyncratic nature of the data and the smaller number of potential TNOs to be processed.
The vetted results of the machine learning process were summarised as a list of detections with a nominal center-time position and velocity in right ascension and declination.
For most detections, if the shift-and-stack step reported multiple `best' velocities, each was initially treated as a separate detection before the later detection-merger steps.

The first step in the linking process was to filter out already-known objects from the detection list.
While this was attempted for the 2020 data, only three previously-known TNOs were visible in the entire 2-field 2020 dataset (2001 BL$_{41}$, 2011 HF$_{103}$, and 2011 JX$_{31}$).
The filtering process turned up significantly more known objects in the 2021 and 2022 detections, all of them being discovered in previous datasets from this program.
This filtering process ensured that known TNOs were not re-linked and saved considerable time in later years. 
Once the linking process was complete, this same filtering code was used to determine which fraction of reported detections corresponded to TNOs with determined orbits; see Table~\ref{tab:detections}.

The next step was to vet the detections visually using the unimplanted star-subtracted HSC data.
These star-subtracted data were produced from the same \verb|processCCD| output as was used for the detection step, but fed through an independent pipeline to produce star-subtracted images.
This independent star subtraction pipeline was previously developed for recovery of \nh\ targets in HSC data obtained in 2016 and 2017 \citep[see][]{Porter2022}.
The images were coregistered to a common sidereal frame and the best 90\%  were stacked to produce a nominal ``star stack''.
The vignetted areas on the edge CCDs and the areas marked as saturated by the LSST pipeline are masked out in the coregistration step.
The effective PSF was extracted for each image as well as for the star-stacked image.
The PSFs were created using a list of 100 stars from Gaia DR3 \citep{Torra2021}, extracting a window of $51\times51$ pixels centered on each star, dividing by the nominal flux of the star based on its Gaia g' magnitude, and then median-combining all the stars to filter out outlier pixels.
For each image, a transformation was calculated from the star stack PSF to the image PSF using the method from \citet{alard1998}, the transformation was applied to the star stack, and the transformed star stack was subtracted from each coregistered image.
This is the same process in principle as applied for the machine learning detections, but using custom Python-based software originally developed for the manual search process before the machine learning was developed.
The dissimilar redundancy of this process would likely not make a difference for fields with much fewer stars, but the extremely high star density of the \nh\ search areas required extra steps to ensure that nominal TNO detections were not just residuals from poor star subtractions.

For each filtered detection, $41\times11$ pixel shift-stacked windows were extracted using the nominal position and velocity information from the detection list.
Each of these shift-stacked windows was then visually vetted, through a process that required the user to either mouse-click on the TNO or press a key to discard the detection.
Typically, 10-30\% of detections were discarded in this step, with the number being much higher for nights with poor seeing which resulted in much less effective star subtraction.
The manual ``click on the TNO'' process was required for earlier iterations of the machine learning process, which tended to only be accurate to within a few pixels, and then retained for resilience as the machine learning process subsequently improved.
The pipeline then used the manual clicked positions as initial conditions to fit the position and flux of the TNO in the shift stack, using the empirical star-based PSF as a model.
After an initial best-fit using downhill simplex as an optimiser, the pixel position and flux of the TNO were fit as a discretely sampled probability distribution function (PDF) using the emcee implementation of a Markov Chain Monte Carlo (MCMC) maximum likelihood technique.
Both the best-fit and MCMC likelihood used a $\chi^2$ estimate based on the sum of the square of the difference between the data and a model-shifted PSF, tuned to the noise of the image, and calibrated on known good detections.
This discretely-sampled pixel and flux cloud was transformed into position and magnitude clouds for the first, middle, and last observations used in the stack.
While this was technically transforming four parameters (mean position RA/Dec and RA/Dec rate) into six, we found it to be best way to capture the motion of the object, since just using two positions resulted in effectively overestimating the motion uncertainty.
This was particularly true for the observations away from opposition, as the motion of the KBOs close to quadrature is very non-linear.
Using three points (instead of two) also effectively encodes the correlation between the points as they were all drawn from the same stacked image. These three clouds were then turned into mean positions and covariance matrices, and the means and covariances saved as astrometry files.

The next step was to fit an initial orbit to this astrometry.
The orbit fitting code used is a derivative of the one described in \citep{Porter2018,Porter2022}, and provides orbit solutions and MCMC clouds that have full planetary perturbations.
To bootstrap the orbit solutions, rough circular orbits were first fit to the data.
These were then immediately used as initial conditions for a second pass that allowed the eccentricity to vary up to 0.1,
followed by a full MCMC solution that allowed eccentricity to vary up to 0.9.
The orbit fitter and MCMC used Cartesian state vectors internally and were thus fully capable of handling parabolic and hyperbolic orbits.
Using Cartesian state vectors was particularly useful for the MCMC, as they represent a fully linearally independent basis, allowing for accurate calculation of their covariance.
However, the observation fields were being shifted at rates optimized for bound, stable objects at similar distances to \nh\ and thus any long-period cometary or interstellar objects would likely have been unrecovered anyways. 
In addition, most of the detections with very low rates of motion were likely star residuals rather than real-but-distant objects, so imposing an effective rate restriction of $e\leq0.9$ filtered out many spurious detections.
Once the initial orbit clouds for the detections were produced, they were then used to create predictions of the positions of the TNO for the images in both the discovery night and in the other observations.
These predictions were produced by propagating the MCMC orbit cloud to the night, determining the RA/Dec of each element of the cloud, and then taking the mean and covariance of the RA/Dec cloud.

At this point, the next iteration of the process begins, a process that is repeated until the orbits are all as well determined as they can be with the dataset.
The predicted positions are first used to find which HSC CCD (if any) the TNO is projected to be on for all the other data being considered, initially just observations close to discovery, then later the full dataset.
The covariance matrices for all the observations on which a TNO is on a CCD are then used to calculate a $3\sigma$ uncertainty ellipse for the TNO on each night.
If the minor axis of the $3\sigma$ uncertainty ellipse was greater than 1000 pixels, then that night was skipped as being too uncertain.
This provided an effective filter for short arcs and initial detections that were projected too far, but still allowed for objects that had a few-month arc (and therefore a well-determined orbit plane) to be projected to data in the previous years, with uncertainty ellipses that might be thousands of pixels long, but only tens of pixels wide.
These ellipses were used to produce World Coordinate System (WCS) projections sized to cover the $3\sigma$ uncertainty and that were rotated to be aligned to the error ellipse, with the horizontal image axis aligned to the major uncertainty axis.
This WCS was then re-centered to the mean predicted locations of the TNO in each image and stacked, to produce a non-linear shift-stack covering the possible locations where the TNO may reside.
Rotating the images to the uncertainty axis saved considerable disk space and CPU time, and meant that most of the shift stack was searchable pixels.
These images were then processed with the same manual inspection process as the initial detections.
The vast majority of detections were just as visible in their discovery night as before (if not moreso), though a small fraction were discarded, as the eccentricity restriction in their initial orbit caused them to no longer appear as a viable detection.
On nights other than the discovery night, the manual process allowed the user to click any source that appeared as the object, or discard that night for that iteration.
This second click would provide a link, and extend the orbit beyond just the discovery night.
In practice, only very bright and obvious TNOs were linked in the first iteration, with more links being attempted as the iterations were repeated and confidence was gained in the detections.
The manual clicks were then used to produce MCMC-fit astrometry just as before, and the astrometry fit to initial orbits with e$<$0.9.
The residuals of all of these initial orbits for an iteration were then visually inspected. 
If a newly added point had significant tension with the existing points for an orbit, it was discarded and the orbit refit.
Orbit clouds and predictions were then generated for all the objects and the process repeated.

Subsequent iterations would allow the observation arcs of the object to be gradually increased, a night or two of data at a time.
Likewise, if an initial link is attempted, and the object does not appear on a third night where it should be visible, that link can be discarded as bad and the object returned to a single night detection.
The detection process is best run on several nights of nearby data simultaneously, as this allows duplication of detection and crosslinking between nights.
The initial duplicate detection checks if any of the not-discovery detections corresponds to known detections for other nights.
Since the machine learning process discussed above can report a detection twice with two different rates, this can mean that up to four detections on two nights can be combined into a single TNO for future iterations.
The next extension is crosslinking, which is only performed on objects that still appear single after the initial manual linking stage, and attempts to fit an orbit for every combination of two single-night detections from different nights.
This is very CPU-intensive, but typically results in a few extra links that were missed by the manual linking process.
Finally, there is a process to merge single night detections with multiple rates, to see if the effective average between their rates is more likely to produce a link than the separate rates by themselves.
This is again not very efficient, but sometimes found additional links that had previously been missed because the original rates reported by KBMOD were incorrect.
After all of these merger or crosslinking steps, the prediction/manual linking/orbit fitting iteration was performed again.
This process was repeated until all possible links had been investigated and the orbits were as good as possible with the data near the discovery.

Once the linking process was finished, the objects with at least two good nights were shifted to a second (but very similar) pipeline that enabled a search of all relevant HSC data back to 2014, as well as include in their orbits any precoveries/recoveries discovered in other datasets, such as the DEEP Survery \citep{Trilling2023}.
This second process was repeated to extend the orbital arcs of the TNO discoveries as much as possible, as well as to find any duplicates with known objects that had not previously been uncovered.
Finally, the resulting astrometry of the second pipeline was exported to MPC submission format and submitted to the MPC.

\section{The Detections \label{sec:detections}}
\begin{figure}
    \includegraphics[width=0.95\textwidth]{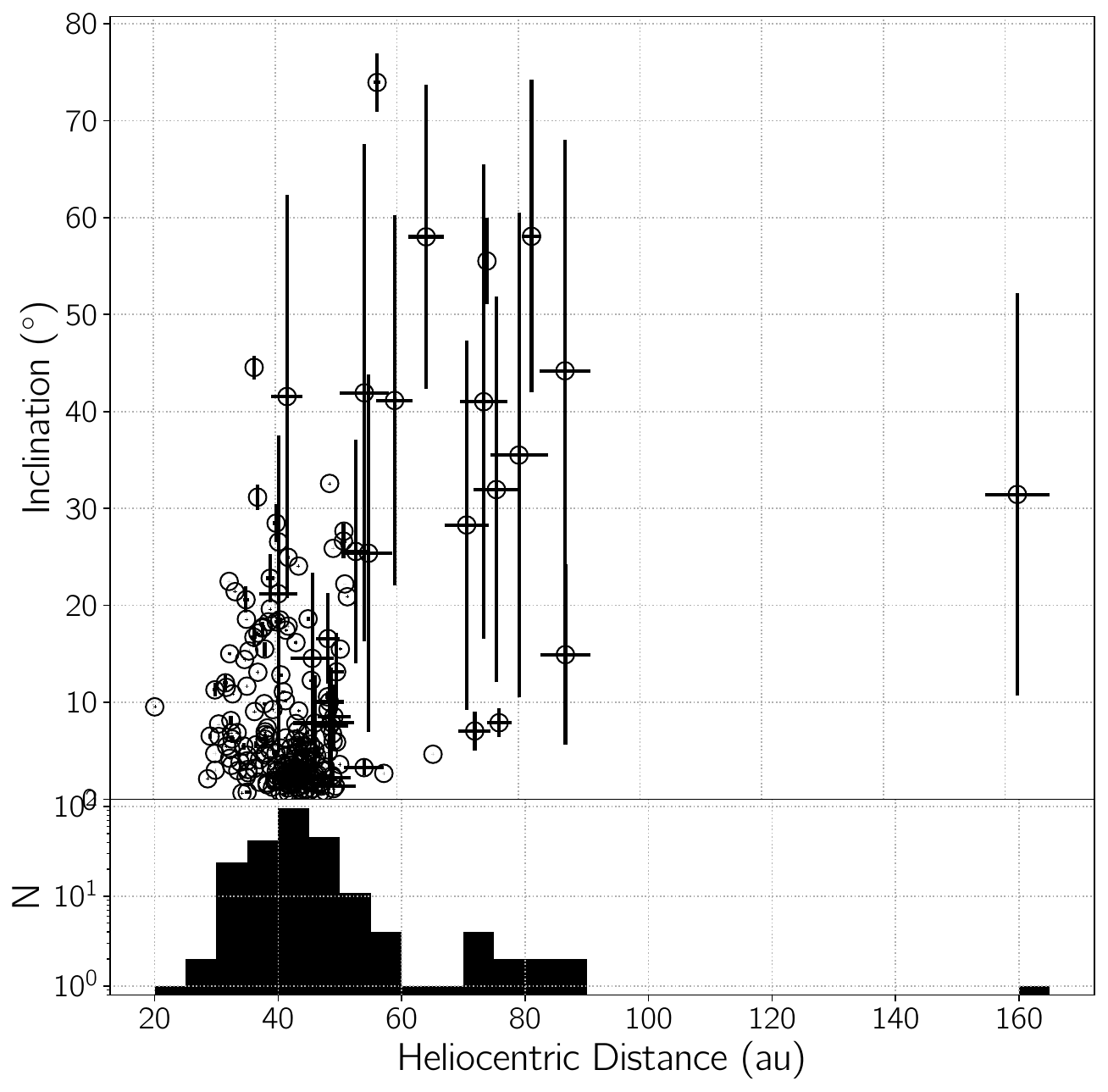}
    \caption{The barycentric inclination and distance of the discoveries we report here. A histogram of the heliocentric distances is shown in the bottom panel, with bin width of 5~au chosen to be similar to the typical uncertainty of the objects with best-fit heliocentric distance $R>70$~au. Uncertainties are $1\sigma$ values from the best-fit orbit clouds (see Section~\ref{sec:linking}).  \label{fig:orbits}}
\end{figure}

We present 239 multi-night detections from our search to date in Tables~\ref{tab:detections} and ~\ref{tab:fulltab2} and Figure~\ref{fig:orbits}. Many of these objects have only a few linked individual detections and of relatively short arcs, and so the fidelity of the derived orbital information is low. However, even for the short-arc detections, the distance at discovery is sufficiently constrained for this discussion.

The orbits are mean barycentric solutions for a cloud of all possible solutions, and since very short arcs have more possible solutions at high inclinations and/or eccentricities, their mean solutions can show exaggerated eccentricity or inclination. True eccentricities and inclinations tend to be biased to lower values than predicted from short arcs, a behavior that is seen by other orbit fitting techniques \citep[see for example][]{Jones2006, Bannister2018}. Forward modelling of simulated ephemerides reveal the tendency for fits to report inflated inclinations for arcs that are a few days in length (see Appendix~\ref{sec:simorbitfit}). However, many of the objects at this point have at least a one-month observational arc, which helps exclude the most inclined solutions. In addition, we excluded solutions with an eccentricity greater than 0.9 for any objects with an arc shorter than two months, again to reduce the number of physically unlikely orbits in the solution cloud. These two constraints together help to produce mean orbits that are not greatly overly inflated by unlikely cases,
and have been proven out in our ability to recover numerous short-arc discoveries in data taken in the following or preceding opposition.

The distance and inclination distributions of our discoveries follow expectations from the known Trans-Neptunian Objects (TNOs) \citep[e.g.,][]{Elliot2005, Bannister2018, Bernardinelli2022}, including the abundance of low-inclination objects between the 3:2 and 2:1 resonances, and the so-called \emph{Kuiper Cliff} -- the well known rapid decrease in detections outside $R\sim55$~au \citep[e.g.,][]{allen2001,Trujillo2001}. Interestingly our detections include an unexpectedly high number of objects at larger heliocentric distances. These objects are the focus of our discussion.

Eleven of our multi-night detections have best-fit heliocentric distances $R>70$~au. Of these \emph{distant} objects, 6 have linkages spanning three or more epochs with three spanning four epochs. The remaining five objects have linkages spanning only two epochs. The high recovery rate lends to the veracity of the detections. Admittedly, our analysis did not include provisions for characterizing the rate of false positive detections and the rate of false linkages that might result. We emphasize however, that three of the five objects with only two epochs of detection have brightnesses $r<26$ which are easy detections in our data suggesting that their lack of more frequent recoveries is merely a reflection of the high stellar density in the field. Indeed, three objects linked across three epochs have brightnesses $r>26.5$ demonstrating that such faint objects can be detected and linked. Furthermore, if the distant detections were indeed driven by false positives, one would expect a pile-up of faint detections near the detection threshold of our survey (typically $r\sim26.5$~mags), but instead, we see that six of the eleven distant objects have brightnesses $<26.0$. 
In general, the rates of motion used for detection by KBMOD are consistent with the rates from final orbits, particularly for the faintest detections, which would only be detectable at nearly the correct rates.

The short arcs (77 days or less) of the distant objects preclude high-accuracy orbit determination (see Table~\ref{tab:detections}). The distance determinations, however, are accurate enough to reveal a valley in the density of detections between $R\sim 55$~au and $R\sim70$~au. To the best of our knowledge, an increase in the number density of objects outside $R\sim70$~au is a feature of the radial distribution that has not been reported by other surveys to similar depths \citep[see for example][]{Smotherman2023}. Given the limited accuracy of the measured orbital elements, simulation of the detailed intrinsic structure implied by our detections would not be useful. Use of the Outer Solar System Origins Survey (OSSOS) Simulator \citep{Lawler2018} however, can still prove informative. A simulation of the pointing history used in this survey and the most up-to-date model of the Outer Solar System planetesimal populations \citep{Kavelaars2021, Petit2023}, which includes a distant belt that is well matched to the OSSOS detections \citep{beaudoin2023} finds that for every 1 detection beyond $\sim70$~au we expect between 72 and 83 detections\footnote{This range is dominated by the uncertainty in the distant population model.} inside that distance. Thus, the 11 distant detections in our survey would imply between $\sim790$ and $\sim900$ detections inside 70~au, which is $3.5-4\times$ larger than we observe. In other words, the 11 distant objects seem to betray the presence of a larger than expected population of bodies beyond $R\sim70$~au. This inference assumes that all of those distant objects are real detections. 
We may have a non-zero false detection rate, though we point out that we have a near 100\% success rate in the follow-up of targets that have been selected for imaging with the HST (22 targets, 1 failed recovery; see Table~\ref{tab:fulltab2}), and the \nh\ spacecraft itself (7 targets). The only two cases of failed follow-up are: a non-detection of a source in LORRI observations; and a non-detection of a source with detection distance $\sim50$~au that was followed up with Gemini imagery. We suspect the latter was due to the short Gemini imaging sequence used which prevented the motion of the source from being detectable. This follow-up track record suggests that the false positive rate cannot be too high. In the extreme circumstance that of the 11 distant sources, the only real objects are those with at least month-long arcs \emph{and} with at least three epochs of detection, we still detect $1.5-1.8\times$ more distant objects than implied by the OSSOS++ model and simulations using our measured detection characteristics. 
Excluding all of our distant detections to bring our survey into compatibility with the OSSOS model would imply an exceptionally high false positive rate. Evidence of a much lower false positive rate comes from the 6 detections with distances $55<r<70$ which have rates of motion only marginally faster than objects with $R>70$~au. Four of those objects have arcs that are months long, with three having arcs years long. Rather, we conclude that many if not most of our distant detections are real. Confirmation of their presence in future observations is highly desirable.



%

\startlongtable
\begin{deluxetable*}{lcccccccc}
        \tablehead{
        \colhead{MPC} &
        \colhead{Arc Length} &
        \colhead{\# Epochs} &
        \colhead{mag (r2)} &
        \colhead{R (au)} &
        \colhead{q (au)} &
        \colhead{a (au)} &
        \colhead{i ($^\circ$)} &
        \colhead{e} \\
        \colhead{} &
        \colhead{ (days)} &
        \colhead{} &
        \colhead{} &
        \colhead{} &
        \colhead{} &
        \colhead{} &
        \colhead{} &
        \colhead{} 
}

\startdata
2017 OW166 & 2242 & 8 & $ 24.52\pm 0.18 $ & $ 43.78\pm 0.00 $ & $ 29.36\pm 0.00 $ & $ 39.50\pm 0.00 $ & $  2.33\pm 0.00 $ & $ 0.26\pm0.00 $ \\
2017 OX166 & 1121 & 8 & $ 26.21\pm 0.26 $ & $ 41.42\pm 0.00 $ & $ 40.84\pm 0.26 $ & $ 44.21\pm 0.01 $ & $  2.43\pm 0.00 $ & $ 0.08\pm0.01 $ \\
2017 OY166 & 1417 & 11 & $ 24.75\pm 0.24 $ & $ 44.20\pm 0.00 $ & $ 40.94\pm 0.00 $ & $ 44.50\pm 0.00 $ & $  2.22\pm 0.00 $ & $ 0.08\pm0.00 $ \\
2020 KA54 & 77 & 6 & $ 26.32\pm 0.21 $ & $ 41.34\pm 0.10 $ & $ 40.20\pm 0.90 $ & $ 47.33\pm 0.59 $ & $ 11.08\pm 0.14 $ & $ 0.15\pm0.01 $ \\
2020 KA55 & 77 & 6 & $ 26.45\pm 0.19 $ & $ 33.85\pm 0.05 $ & $ 33.79\pm 0.08 $ & $ 36.82\pm 0.46 $ & $  2.95\pm 0.01 $ & $ 0.08\pm0.01 $ \\
2022 LU16 & 28 & 2 & $ 26.12\pm 0.52 $ & $ 71.48\pm 3.65 $ & $ 19.78\pm18.61 $ & $ 84.48\pm54.48 $ & $ 28.29\pm19.05 $ & $ 0.82\pm0.16 $ \\
2022 LT16 & 28 & 3 & $ 25.79\pm 1.29 $ & $ 72.81\pm 2.64 $ & $ 29.30\pm17.25 $ & $ 105.16\pm76.29 $ & $  7.04\pm 2.01 $ & $ 0.68\pm0.24 $ \\
2022 LR15 & 28 & 2 & $ 26.32\pm 0.35 $ & $ 74.28\pm 3.90 $ & $ 21.73\pm18.89 $ & $ 86.64\pm52.83 $ & $ 41.00\pm24.45 $ & $ 0.80\pm0.15 $ \\
2020 KX11 & 77 & 5 & $ 26.09\pm 0.56 $ & $ 74.81\pm 0.29 $ & $  4.11\pm 0.35 $ & $ 110.07\pm42.70 $ & $ 55.52\pm 4.45 $ & $ 0.96\pm0.01 $ \\
2022 LY15 & 27 & 2 & $ 25.64\pm 1.00 $ & $ 76.37\pm 3.80 $ & $ 23.13\pm19.94 $ & $ 116.38\pm108.28 $ & $ 31.94\pm19.89 $ & $ 0.80\pm0.17 $ \\
2022 LQ16 & 29 & 4 & $ 26.43\pm 0.48 $ & $ 76.81\pm 2.04 $ & $ 44.22\pm16.60 $ & $ 147.52\pm123.79 $ & $  7.89\pm 1.47 $ & $ 0.60\pm0.25 $ \\
2020 PD95 & 3 & 2 & $ 25.39\pm 1.72 $ & $ 80.08\pm 4.74 $ & $ 22.11\pm20.41 $ & $ 98.18\pm66.82 $ & $ 35.50\pm24.97 $ & $ 0.83\pm0.14 $ \\
2022 LX15 & 28 & 3 & $ 25.97\pm 0.52 $ & $ 82.15\pm 1.56 $ & $  9.13\pm 2.09 $ & $ 83.13\pm53.49 $ & $ 58.07\pm16.12 $ & $ 0.86\pm0.08 $ \\
2022 LS16 & 29 & 2 & $ 25.59\pm 0.94 $ & $ 87.65\pm 4.20 $ & $ 27.74\pm21.95 $ & $ 122.65\pm97.52 $ & $ 44.19\pm23.82 $ & $ 0.79\pm0.16 $ \\
2020 MJ53 & 4 & 2 & $ 26.25\pm 0.44 $ & $ 87.72\pm 4.06 $ & $ 23.91\pm24.35 $ & $ 114.66\pm89.28 $ & $ 14.91\pm 9.30 $ & $ 0.84\pm0.15 $ \\
2020 MK53 & 4 & 2 & $ 25.99\pm 0.49 $ & $ 162.03\pm 5.35 $ & $ 46.54\pm45.03 $ & $ 212.81\pm161.42 $ & $ 31.44\pm20.77 $ & $ 0.83\pm0.15 $ \\
\enddata
\tablecomments{
Barycentric orbital elements of the discovered objects. Short example of the list of TNOs discovered in our 2020, 2021, and 2022 Subaru-HSC searches.
These discoveries have been submitted to the Minor Planet Center. 
See Section~\ref{sec:linking} for details on the orbit fitting. The full list is presented in the Appendix. The point clouds generated as part of the orbit fitting are available from the authors upon request. For convenience we have included the 11 objects with $R>70$~au. \label{tab:detections}
}
\end{deluxetable*}

\section{Discussion \label{sec:discussion}}

Our use of Subaru Telescope, which was needed to search the wide area where candidates might be located, resulted in a sample of 239 TNOs being discovered as of December 2023, including those TNOs that were suitable for observations from \nh. 
The observing strategy and search technique we employed enabled precise determination of the detection and tracking efficiency of the sample resulting in a well-characterized sample of TNO orbits. Comparison of our sample with outputs from the OSSOS Survey Simulator reveals a $\sim3.5-4\times$ larger-than-expected sample of objects at distances $R\gtrsim70$~au.


As noted in \citet{Shannon2023}, there is a growing body of evidence hinting at the presence of a yet-to-be-discovered mass of distant planetesimals. This evidence includes the dust impact rate measurements of the Student Dust Counter on-board \nh\ which are not decreasing at the rate expected given the known TNOs \citep{Doner2024}, the serendipitous occultation detections \citep[e.g.,][]{Schlichting2012, Arimatsu2019} which seem too improbable given the known extent of the dynamically excited Kuiper Belt, and the report of a few distant TNOs detected in archival HST imagery \citep{Fuentes2010} all of which which appear to be incompatible with the observed luminosity function reported from other ground-based surveys \citep{Fraser2014}.  As Shannon et al. point out, models that include a more massive distant population with either or some combination of decreasing maximum size distribution with increasing semi-major axis, or increasing surface density of material, can reconcile the occultation detections with the Pluto-Charon cratering record \citep{Singer2019}. Indeed, at least in principle, all of these results may be reconciled if there is a population of planetesimals beyond the main Kuiper Belt that has yet to be recognized.

Some rough constraints on the possible distant population come from past TNO surveys. The deepest published wide area surveys with good distance measurements for their sources have a brightness limit of $r\sim25.5$~mag \citep{Bannister2018} and do not report the detection of a substantial distant population that could explain the abundance of objects at distances $\sim70$~au that we report here. If the distant population was in a belt form like that of the classical TNOs, it must be the case that the majority of objects must be fainter than $r=25.5$, or $H\sim7$. These are significantly smaller than the largest cold classical objects, which have $H\sim5.0$ \citep{Kavelaars2021}, and is consistent with the suggestion by Shannon that the size distribution of an outer population is likely to consist of smaller bodies than found in the main Kuiper Belt. With the use of the OSSOS Survey Simulator \citep{Lawler2018}, we have found that the 11 detections with distance $R>70$~au that we report are consistent with a belt of material in a 5~au wide ring centered at 70~au with a surface density of 6.2 objects per square au with $H_{r_2}<8.66$. This estimate assumes the objects have a steep size distribution like that of the hot classical TNOs \citep{Petit2023} of the form $N(<r_2)\propto10^{\alpha r_2}$ with $\alpha=1.1$ for objects with $H_{r_2}<7.5$ and $\alpha=0.3$ for objects with $H_{r_2}>7.5$ and are on circular orbits with inclinations distributed as a sin-normal with width of $12^\circ$ \citep{PeltierACM2023}. This density increases to 20.5 objects per square au if the ring is instead centered at 80~au. These density estimates suggest a mass that is comparable with the mass of the known excited TNO population but these estimates are highly uncertain, and so provide only a sense of the scale of the required population, as the orbital and size distributions of such a population remain mostly unconstrained. The orbital elements of our distant detections imply at least moderate inclinations for those bodies, and so we do not expect the underlying population to be dynamically cold. This is consistent with the moderate ecliptic latitudes of the detections of serendipitous stellar occultations reported by \citet{Schlichting2012}.

\begin{figure}
    \includegraphics[width=0.95\textwidth]{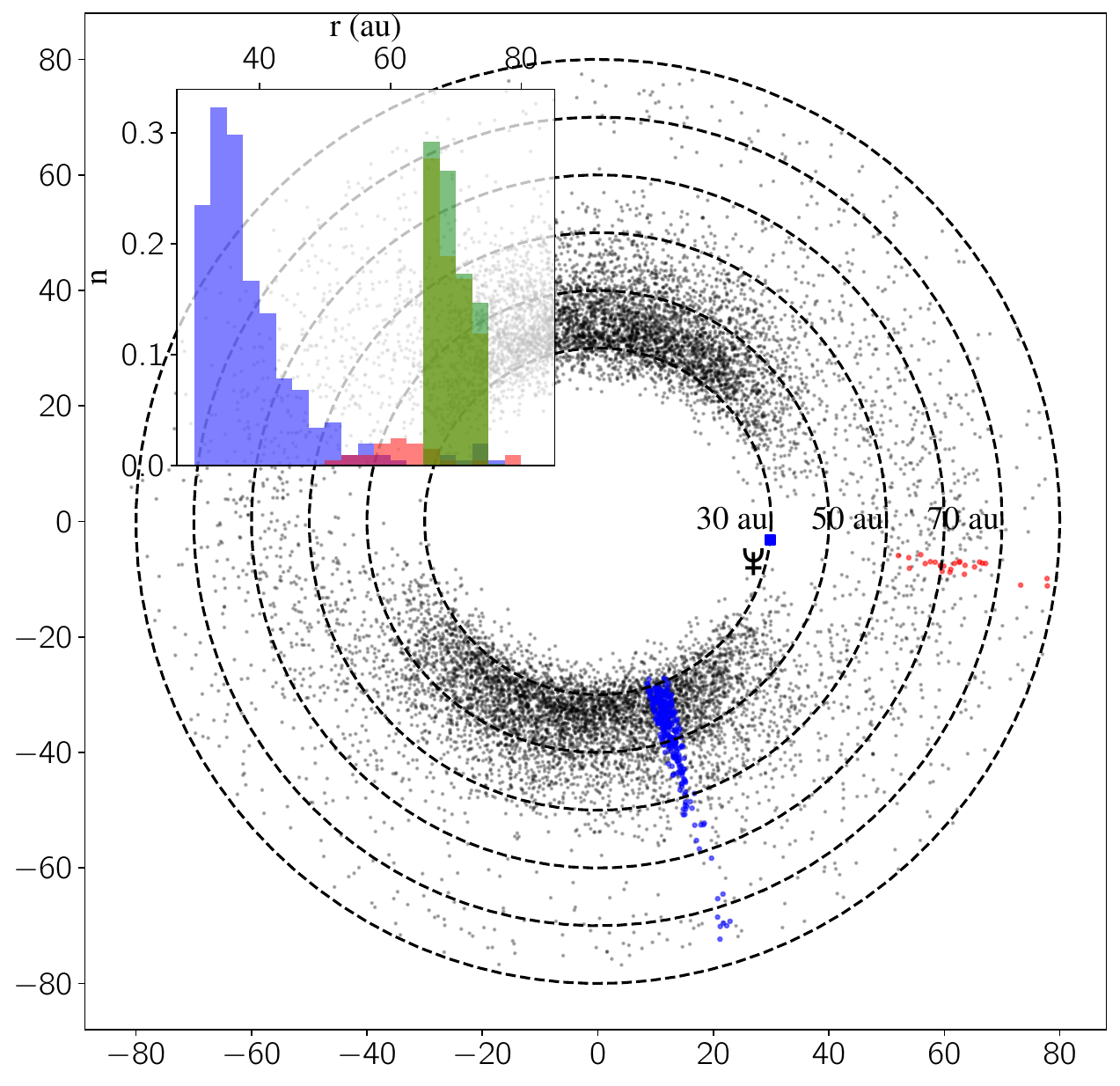}
    \caption{The spatial distribution of the model 5-2 resonant objects in the OSSOS++ model realized on Julian Date 2459874.5. The position of Neptune is shown by the blue square. Only those objects with r$<27$ are shown. The realization is $\sim8\times$ over sampled. Objects seen by the \nh\ and DEEP pointings are shown in blue and red, respectively, with color-matched histograms of each sample shown in the inset. As well, the inset also shows an example of a radial distribution of discoveries that would be discovered from an annulus model in the \nh\ (orange) and DEEP (green) pointings. The example is of an annulus centered at 70~au. The annuli histograms have been rescaled for clarity. Annuli centered at other distances merely move the histogram of simulated detections inward or outward accordingly.  This figure demonstrates how strongly survey pointing can influence the density and distribution of detectable resonant objects, an effect that is not present in distributions that are longitudinally symmetric. \label{fig:5to2}}
\end{figure}

While the presence of an undiscovered belt is compatible with the distribution of distant objects we report, it is incompatible with the results of the DECam Ecliptic Exploration Project \citep[DEEP][]{Trilling2023}. For an example, we show a simulated distribution of detections for both DEEP and this survey, for a belt or annulus of distant bodies in Figure~\ref{fig:5to2}. DEEP reports a similar limiting magnitude and survey area and utilizes very similar search techniques as this survey does, but detected only 1 object with distance $R>60$~au \citep{Napier2023, Bernardinelli2023, Smotherman2023}.  The density of detections of objects at $R>60$~au in DEEP is roughly an order of magnitude lower than we report here, where the density of objects should be the same for both surveys if the population of distant bodies is in a longitudinally symmetric distribution such as in the annulus model. This result implies  either that DEEP are unexpectedly insensitive to slow-moving bodies, or, more likely, that if there is a large population of distant bodies that their apparent sky density is low at the DEEP survey pointing locations.

The differences in the observed distributions of distant objects between DEEP and what we report here may be caused by longitudinal variations in the radial distribution of objects in distant mean-motion resonances. To demonstrate how this might be the case, we present a toy model realization of the 5:2 resonant population in the OSSOS++ model. This population uses a uniform semi-major axis distribution between $54.75\leq a\leq 55.75$~au, a Gaussian distribution in eccentricity with mean of 0.43 and standard deviation of 0.03 limited to $0.2 \leq e \leq 0.5$, a sin-normal inclination distribution with width of $20^\circ$ limited to $i\leq 45^\circ$, and a triangularly distributed resonant amplitude with peak at $75^\circ$ that goes to zero at $0^\circ$ and $180^\circ$. We adopted the size distribution for the hot classical population reported by \citet{Petit2023}.
A realization of the aforementioned 5:2 model was generated with the OSSOS Survey Simulator \citep{Lawler2018}, drawing objects randomly through the Solar System until 1,000 total objects were drawn between the \nh\ Survey, and the DEEP pointing. The former used the exact pointing history that we report in Table~1, and the latter was a single epoch pointing at 23:32:00, -04:00:00 with an area of 7.9 square degrees. Those drawn objects with magnitude $r<27$ were recorded, and are presented in Figure~\ref{fig:5to2}. 
This simple realization broadly reflects the notable properties of the real detections we report here, including an absence of objects near $R\sim65$~au with a smattering of detections at further distances. 
Such a longitude dependent distribution is also consistent with the lack of detections in the DEEP sample beyond $R\sim70$~au. 

We emphasize that our 5:2 population is a toy model that uses a single resonance with a relatively poorly constrained eccentricity distribution \citep{MatthewsThesis2019,Crompvoets2022} to demonstrate the influence of pointing longitude on the observed sample. The true Solar System includes many other resonant populations.  Confirmation that the distribution we have observed is driven by a heretofore previously unrecognized massive resonant population awaits significant future modeling efforts. Though we point out that a simulation of the DEEP and \nh\ pointings shown in Figure~\ref{fig:5to2} that utilizes the full OSSOS++ model  -- which includes an up-to-date estimate of of all distant resonances for which at least one object was detected \citep{Crompvoets2022} -- presents a similar lack of distant detections compared to more proximate TNO populations. 

Whatever the intrinsic dynamical structure distant ($R\gtrsim70$~au) objects occupy -- disk, annulus, resonant population, etc. -- the discovery rate in the survey we present here implies more mass in that distant population than previously recognized. A larger than expected population of distant TNOs can help explain the unexpected and consistently high Student Dust Counter impact rates, and the serendipitous occultations detected by \citet{Schlichting2012} and \citet{Arimatsu2019}, both of which appear somewhat at odds with the small known mass of TNOs at the required distances. With an increased mass, these observations may not be at odds after all. 

The work we have presented here is not an in-depth analysis of the broad and varied range of potential forms a distant population could take. Any detailed analysis will be hampered by the limited fidelity of the orbits we report which are simply insufficient to provide strong constraint on the orbital distributions the distant objects may take.  Rather, we have limited our discussion mainly to our methods of discovery and stick to the main point -- there appears to be more distant bodies than previously recognized.

Constraints on the orbital structure of the distant population remain weak, primarily as a result of the insufficient depth limits of surveys which have provided the strongest orbital constraints. A survey with depth limits similar to what we report here, but over a wider field and at different ecliptic longitudes is necessary to confirm the presence of this distant population, and to reveal its intrinsic orbital structure. The wait for such a survey should not be too long however, as combined ongoing surveys like the DEEP survey and the Classical and Large-A Solar System Survey \citep[CLASSY][]{FraserACM2023}, the pencil beam survey with the James Webb Space Telescope \citep{Eduardo2023, Morgan2023}, and the efforts of this team which include on-going observations and the application of additional search techniques to the data we present here \citep[e.g.,][]{Yoshida2024}, should achieve sufficient depth, coverage, and spread in pointings to detect large numbers of these distant bodies, and test the findings we report here.

\section{Summary  \label{sec:conclusions}}
We report the discovery of 239 new TNOs discovered with the Subaru Telescope and the Hyper SuprimeCam as part of the \nh\ distant search.  Beyond the \emph{Kuiper Cliff}, these objects show an unexpectedly high fraction of distant $R\gtrsim70$~au hinting at a previously unrecognized large population of distant bodies. The fidelity of the orbits for these distant objects is too low to provide meaningful constraints on the intrinsic orbital structure of the distant population, which is compatible with a homogeneous distribution (disk or annulus) or a large population of resonant bodies. The inferred mass however, could be as high as the known excited Kuiper Belt. Confirmation of this signal is required, and will likely come from ongoing surveys, including this one. 

\bibliographystyle{aastex631}
\bibliographystyle{aasjournal}
\bibliography{NHKBOs_ADS_LIBRARY_v1}

\software{ parallel \citep{Tange2018}, numpy \citep{Harris2020}, matplotlib \citep{Hunter2007}, TRailed Imaging in Python, \citep{Fraser2016}, astropy \citep{Astropy2013, Astropy2018}, SExtractor \citep{hihi}}

\section{Acknowledgements}

The authors wish to recognize and acknowledge the very significant cultural role and reverence that the summit of Maunakea has always had within the indigenous Hawaiian community.  We are most fortunate to have the opportunity to conduct observations from this mountain.

This research is based on data collected at the 8.2m Subaru Telescope (\url{https://subarutelescope.org/en/}) using its prime focus camera (the Hyper Suprime-Cam), operated by the National Astronomical Observatory of Japan. The Subaru Telescope provided the main source of data for this work. We would like to thank the director and the TAC for working with us to accommodate our complex scheduling requirements.  

This research is based on observations from the programs GS-2019B-FT-102, GN-2020A-DD-103, and GS-2020A-DD-103 obtained at the international Gemini Observatory, a program of NSF’s NOIRLab, which is managed by the Association of Universities for Research in Astronomy (AURA) under a cooperative agreement with the National Science Foundation on behalf of the Gemini Observatory partnership: the National Science Foundation (United States), National Research Council (Canada), Agencia Nacional de Investigaci\'{o}n y Desarrollo (Chile), Ministerio de Ciencia, Tecnolog\'{i}a e Innovaci\'{o}n (Argentina), Minist\'{e}rio da Ci\^{e}ncia, Tecnologia, Inova\c{c}\~{o}es e Comunica\c{c}\~{o}es (Brazil), and Korea Astronomy and Space Science Institute (Republic of Korea).

This research used the Canadian Astronomy Data Centre facilities operated by the National Research Council of Canada with the support of the Canadian Space Agency. This work also made use of the Gemini Observatory Archive, NASA’s Astrophysics Data System Bibliographic Services, and the JPL HORIZONS web interface (\url{https://ssd.jpl.nasa.gov/horizons.cgi}).

\nh\ scientists Stern, Benecchi, Buie, Fraser, Lauer, Kavelaars, Peltier, Porter, and Verbiscer thank NASA's New Horizons Mission (NASW-02008) for funding.

This work made use of the Pan-STARRS1 Surveys (PS1) and the PS1 public science archive, which have been made possible through contributions by the Institute for Astronomy, the University of Hawaii, the Pan-STARRS Project Office, the Max-Planck Society and its participating institutes, the Max Planck Institute for Astronomy, Heidelberg and the Max Planck Institute for Extraterrestrial Physics, Garching, The Johns Hopkins University, Durham University, the University of Edinburgh, the Queen's University Belfast, the Harvard-Smithsonian Center for Astrophysics, the Las Cumbres Observatory Global Telescope Network Incorporated, the National Central University of Taiwan, the Space Telescope Science Institute, the National Aeronautics and Space Administration under Grant No. NNX08AR22G issued through the Planetary Science Division of the NASA Science Mission Directorate, the National Science Foundation Grant No. AST-1238877, the University of Maryland, Eotvos Lorand University (ELTE), the Los Alamos National Laboratory, and the Gordon and Betty Moore Foundation.

This paper makes use of LSST Science Pipelines software developed by the Vera C. Rubin Observatory. We thank the Rubin Observatory for making their code available as free software at \url{https://pipelines.lsst.io}.

We also thank the former and current directors of the Subaru Telescope, Michitoshi
Yoshida and Satoshi
Miyazaki, for opening the Subaru Telescope to collaborate with the New
Horizons mission. And we thank the TAC and Subaru staff for working with us to
accommodate our complex scheduling requirements.

Finally, W. C. Fraser would like to thank B. C. Pontus and D. Ragozzine for their insightful comments.

\appendix
\restartappendixnumbering

\section{Full Detection List}

\startlongtable

\startlongtable
\begin{deluxetable*}{lcccccccccccccc}
\tablecaption{The fitted barycentric orbital elements of the TNO discoveries reported in this paper. Objects are sorted in order of increase heliocentric distance. \label{tab:fulltab2}}
\tablehead{
\colhead{MPC} & \colhead{Arc} &  \colhead{Epochs} & \colhead{mag} & \colhead{$\sigma_{mag}$} & \colhead{R} &  \colhead{$\sigma_R$} & \colhead{q} &  \colhead{$\sigma_q$} & \colhead{a} & \colhead{$\sigma_a$} & \colhead{inc} & \colhead{$\sigma_{inc}$} & \colhead{e} & \colhead{$\sigma_e$} \\
 & \colhead{(day)} & \colhead{\#} & & & \colhead{(au)} & \colhead{(au)} & \colhead{(au)} & \colhead{(au)} &  \colhead{(au)} & \colhead{(au)} & \colhead{(deg)} & \colhead{(deg)} & & 
}
\startdata
2017 OW166 & 2242 & 8 &  24.52 &  0.18  &  43.78 & 0.00  &  29.36 &  0.00  &  39.50 &  0.00  &   2.33 &  0.00  &  0.26 & 0.00  \\
2017 OX166 & 1121 & 8 &  26.21 &  0.26  &  41.42 & 0.00  &  40.84 &  0.26  &  44.21 &  0.01  &   2.43 &  0.00  &  0.08 & 0.01  \\
2017 OY166 & 1417 & 11 &  24.75 &  0.24  &  44.20 & 0.00  &  40.94 &  0.00  &  44.50 &  0.00  &   2.22 &  0.00  &  0.08 & 0.00  \\
2020 KA54 & 77 & 6 &  26.32 &  0.21  &  41.34 & 0.10  &  40.20 &  0.90  &  47.33 &  0.59  &  11.08 &  0.14  &  0.15 & 0.01  \\
2020 KA55 & 77 & 6 &  26.45 &  0.19  &  33.85 & 0.05  &  33.79 &  0.08  &  36.82 &  0.46  &   2.95 &  0.01  &  0.08 & 0.01  \\
2020 KA56 & 79 & 8 &  24.81 &  0.21  &  38.37 & 0.04  &  32.11 &  0.32  &  39.69 &  0.67  &   4.57 &  0.03  &  0.19 & 0.02  \\
2020 KB54 & 76 & 8 &  25.53 &  0.27  &  44.47 & 0.07  &  41.42 &  0.63  &  44.04 &  0.75  &   2.73 &  0.01  &  0.06 & 0.02  \\
2020 KB55 & 77 & 6 &  26.06 &  0.18  &  45.40 & 0.10  &  44.80 &  0.58  &  47.74 &  0.83  &  18.61 &  0.26  &  0.07 & 0.02  \\
2020 KB56 & 79 & 7 &  25.93 &  0.25  &  43.39 & 0.11  &  42.83 &  0.28  &  143.42 &  6.11  &   7.80 &  0.09  &  0.70 & 0.01  \\
2020 KC53 & 78 & 9 &  25.84 &  0.25  &  44.42 & 0.09  &  42.25 &  1.52  &  43.96 &  0.63  &   2.39 &  0.01  &  0.04 & 0.03  \\
2020 KC54 & 79 & 7 &  24.73 &  0.29  &  42.82 & 0.05  &  41.08 &  1.19  &  42.54 &  0.34  &   2.59 &  0.00  &  0.03 & 0.02  \\
2020 KC55 & 77 & 6 &  26.08 &  0.25  &  44.17 & 0.09  &  38.64 &  2.04  &  43.91 &  0.38  &   2.70 &  0.00  &  0.12 & 0.04  \\
2020 KC56 & 79 & 5 &  25.17 &  0.25  &  38.22 & 0.12  &  36.80 &  1.33  &  47.91 &  1.31  &   9.86 &  0.15  &  0.23 & 0.04  \\
2020 KD53 & 467 & 8 &  25.37 &  0.19  &  35.67 & 0.00  &  34.86 &  0.02  &  96.98 &  0.17  &  15.29 &  0.00  &  0.64 & 0.00  \\
2020 KD54* & 423 & 8 &  26.28 &  0.25  &  42.14 & 0.00  &  37.35 &  0.03  &  66.82 &  0.09  &  17.85 &  0.00  &  0.44 & 0.00  \\
2020 KD55 & 76 & 6 &  25.95 &  0.24  &  32.53 & 0.06  &  31.74 &  0.42  &  36.48 &  0.68  &  15.00 &  0.15  &  0.13 & 0.03  \\
2020 KD56 & 27 & 4 &  25.60 &  0.22  &  37.92 & 0.28  &  32.71 &  3.83  &  113.54 & 81.73  &  17.65 &  0.65  &  0.64 & 0.20  \\
2020 KE53 & 406 & 9 &  26.37 &  0.21  &  39.17 & 0.00  &  34.48 &  0.20  &  39.33 &  0.05  &   5.06 &  0.00  &  0.12 & 0.01  \\
2020 KE54 & 80 & 6 &  23.03 &  0.21  &  33.41 & 0.01  &  31.03 &  0.11  &  39.59 &  0.21  &  21.41 &  0.05  &  0.22 & 0.01  \\
2020 KE55 & 77 & 4 &  26.43 &  0.21  &  42.93 & 0.19  &  39.35 &  3.01  &  47.99 &  2.64  &   2.98 &  0.04  &  0.18 & 0.09  \\
2020 KE56 & 75 & 5 &  24.85 &  0.21  &  38.86 & 0.05  &  33.08 &  0.32  &  39.01 &  0.59  &   6.52 &  0.04  &  0.15 & 0.02  \\
2020 KF53 & 5 & 3 &  26.06 &  0.17  &  49.90 & 3.30  &  16.73 & 12.19  &  61.10 & 36.58  &   1.35 &  0.10  &  0.72 & 0.23  \\
2020 KF54 & 80 & 7 &  24.96 &  0.20  &  40.17 & 0.04  &  38.42 &  0.60  &  43.79 &  0.26  &   2.91 &  0.00  &  0.12 & 0.01  \\
2020 KF55 & 75 & 5 &  25.99 &  0.33  &  40.93 & 0.10  &  36.36 &  1.09  &  46.18 &  1.96  &  12.82 &  0.18  &  0.21 & 0.06  \\
2020 KF56 & 78 & 5 &  25.97 &  0.23  &  41.68 & 0.10  &  40.86 &  0.70  &  42.84 &  0.89  &  10.16 &  0.14  &  0.05 & 0.03  \\
2020 KG53 & 5 & 3 &  25.61 &  0.25  &  49.54 & 2.89  &  17.70 & 11.33  &  60.64 & 34.83  &   2.22 &  0.66  &  0.70 & 0.23  \\
2020 KG54* & 462 & 8 &  25.66 &  0.22  &  32.42 & 0.00  &  32.19 &  0.00  &  78.33 &  0.02  &  22.47 &  0.00  &  0.59 & 0.00  \\
2020 KG55 & 24 & 4 &  26.14 &  0.20  &  38.27 & 0.35  &  28.96 &  5.39  &  79.23 & 55.58  &  15.49 &  0.75  &  0.53 & 0.27  \\
2020 KG56 & 76 & 9 &  26.13 &  0.21  &  35.22 & 0.05  &  31.88 &  0.40  &  47.44 &  1.38  &   2.29 &  0.00  &  0.33 & 0.03  \\
2020 KH42* & 400 & 8 &  22.36 &  2.64  &  49.53 & 0.00  &  37.32 &  0.01  &  77.55 &  0.05  &  25.88 &  0.00  &  0.52 & 0.00  \\
2020 KH54 & 80 & 7 &  25.18 &  0.26  &  44.10 & 0.07  &  38.19 &  0.69  &  44.46 &  0.96  &   3.11 &  0.01  &  0.14 & 0.03  \\
2020 KH55 & 27 & 4 &  25.73 &  0.21  &  44.06 & 0.37  &  28.24 &  6.52  &  82.60 & 67.31  &   3.77 &  0.12  &  0.53 & 0.29  \\
2020 KH56* & 426 & 9 &  25.33 &  0.34  &  35.53 & 0.00  &  32.63 &  0.14  &  55.68 &  0.36  &   3.06 &  0.00  &  0.41 & 0.01  \\
2020 KJ54 & 77 & 5 &  25.42 &  0.17  &  43.63 & 0.08  &  40.85 &  1.65  &  42.81 &  0.54  &   2.45 &  0.00  &  0.05 & 0.03  \\
2020 KJ55 & 27 & 4 &  25.89 &  0.24  &  44.86 & 0.39  &  30.00 &  6.37  &  68.07 & 33.60  &   2.06 &  0.00  &  0.50 & 0.28  \\
2020 KJ56* & 371 & 9 &  22.60 &  0.30  &  42.80 & 0.00  &  42.79 &  0.01  &  44.09 &  0.01  &   2.73 &  0.00  &  0.03 & 0.00  \\
2020 KK54* & 155 & 7 &  25.92 &  0.20  &  49.53 & 0.00  &  37.25 &  0.26  &  43.97 &  0.15  &   5.99 &  0.00  &  0.15 & 0.01  \\
2020 KK55* & 369 & 7 &  26.43 &  0.24  &  34.22 & 0.00  &  33.75 &  0.09  &  76.21 &  0.35  &   3.83 &  0.00  &  0.56 & 0.00  \\
2020 KL53* & 388 & 8 &  24.14 &  0.37  &  44.59 & 0.00  &  42.46 &  0.04  &  43.59 &  0.01  &   2.41 &  0.00  &  0.03 & 0.00  \\
2020 KL54 & 77 & 6 &  25.89 &  0.23  &  43.67 & 0.11  &  38.67 &  2.60  &  43.77 &  0.36  &   7.09 &  0.10  &  0.12 & 0.05  \\
2020 KL55 & 75 & 5 &  26.07 &  0.40  &  46.56 & 0.17  &  23.63 &  1.53  &  176.05 & 63.48  &   2.51 &  0.03  &  0.86 & 0.06  \\
2020 KM53 & 388 & 6 &  25.14 &  0.28  &  44.06 & 0.00  &  41.93 &  0.64  &  44.53 &  0.05  &   1.96 &  0.00  &  0.06 & 0.02  \\
2020 KM54 & 76 & 6 &  26.09 &  0.24  &  41.73 & 0.10  &  39.17 &  1.47  &  45.89 &  0.58  &   6.33 &  0.08  &  0.15 & 0.02  \\
2020 KM55 & 77 & 6 &  25.70 &  0.19  &  44.41 & 0.09  &  40.90 &  1.81  &  45.88 &  0.46  &   2.67 &  0.01  &  0.11 & 0.03  \\
2020 KN53 & 78 & 10 &  24.76 &  0.19  &  39.63 & 0.04  &  38.73 &  0.40  &  42.96 &  0.25  &   9.25 &  0.04  &  0.10 & 0.01  \\
2020 KN54 & 78 & 7 &  24.27 &  0.21  &  20.19 & 0.01  &  11.29 &  0.05  &  16.83 &  0.01  &   9.53 &  0.02  &  0.33 & 0.00  \\
2020 KN55 & 77 & 6 &  24.22 &  0.17  &  40.56 & 0.03  &  37.16 &  0.57  &  41.13 &  0.13  &  26.54 &  0.11  &  0.10 & 0.01  \\
2020 KO11* & 154 & 8 &  25.24 &  0.25  &  49.42 & 0.00  &  38.54 &  0.11  &  44.75 &  0.05  &   6.75 &  0.00  &  0.14 & 0.00  \\
2020 KO53 & 77 & 6 &  25.32 &  0.14  &  45.93 & 0.09  &  42.64 &  1.16  &  48.28 &  1.20  &  12.26 &  0.14  &  0.12 & 0.05  \\
2020 KO54 & 29 & 5 &  26.15 &  0.44  &  29.30 & 0.25  &  23.59 &  3.98  &  54.51 & 23.15  &   6.51 &  0.26  &  0.53 & 0.23  \\
2020 KO55 & 77 & 6 &  25.86 &  0.19  &  44.94 & 0.09  &  37.26 &  1.59  &  47.74 &  0.45  &   5.47 &  0.05  &  0.22 & 0.03  \\
2020 KP11* & 1202 & 27 &  23.74 &  2.57  &  51.45 & 0.00  &  38.10 &  0.00  &  45.47 &  0.00  &  22.20 &  0.00  &  0.16 & 0.00  \\
2020 KP53 & 77 & 6 &  26.03 &  0.22  &  45.87 & 0.11  &  38.66 &  0.76  &  46.11 &  1.53  &   4.37 &  0.05  &  0.16 & 0.04  \\
2020 KP54 & 77 & 7 &  26.17 &  0.24  &  41.11 & 0.09  &  38.63 &  1.08  &  43.02 &  1.12  &   2.93 &  0.02  &  0.10 & 0.05  \\
2020 KP55* & 366 & 8 &  25.04 &  0.27  &  32.82 & 0.00  &  30.31 &  0.12  &  47.87 &  0.18  &   3.49 &  0.00  &  0.37 & 0.00  \\
2020 KQ11* & 156 & 8 &  23.20 &  0.35  &  48.23 & 0.00  &  39.56 &  0.05  &  44.55 &  0.01  &   3.32 &  0.00  &  0.11 & 0.00  \\
2020 KQ54 & 77 & 7 &  26.36 &  0.30  &  42.47 & 0.11  &  41.85 &  0.61  &  44.45 &  0.88  &   2.89 &  0.03  &  0.06 & 0.02  \\
2020 KQ55 & 77 & 6 &  26.09 &  0.21  &  43.86 & 0.09  &  42.39 &  1.24  &  45.59 &  0.67  &   9.11 &  0.11  &  0.07 & 0.02  \\
2020 KR11* & 1771 & 15 &  22.19 &  2.73  &  50.09 & 0.00  &  42.67 &  0.00  &  46.38 &  0.00  &   5.89 &  0.00  &  0.08 & 0.00  \\
2020 KR54 & 77 & 7 &  25.84 &  0.22  &  41.06 & 0.08  &  40.79 &  0.27  &  44.07 &  0.70  &   2.04 &  0.00  &  0.08 & 0.01  \\
2020 KR55 & 80 & 7 &  25.95 &  0.17  &  38.89 & 0.06  &  38.77 &  0.11  &  42.60 &  0.46  &  18.30 &  0.14  &  0.09 & 0.01  \\
2020 KS11* & 792 & 9 &  26.42 &  0.30  &  57.86 & 0.00  &  34.92 &  0.02  &  99.86 &  0.19  &   2.67 &  0.00  &  0.65 & 0.00  \\
2020 KS53* & 370 & 10 &  24.74 &  0.25  &  45.82 & 0.00  &  43.49 &  0.06  &  44.69 &  0.02  &   5.67 &  0.00  &  0.03 & 0.00  \\
2020 KS54 & 79 & 7 &  25.92 &  0.33  &  40.77 & 0.09  &  38.14 &  1.67  &  42.01 &  0.39  &  18.50 &  0.23  &  0.09 & 0.03  \\
2020 KS55 & 80 & 7 &  25.74 &  0.31  &  42.14 & 0.06  &  37.20 &  0.53  &  59.75 &  1.96  &  24.94 &  0.18  &  0.38 & 0.03  \\
2020 KT11* & 1391 & 14 &  24.40 &  2.61  &  51.30 & 0.00  &  33.74 &  0.00  &  42.90 &  0.00  &  27.65 &  0.00  &  0.21 & 0.00  \\
2020 KT53 & 75 & 5 &  25.22 &  0.15  &  45.61 & 0.09  &  40.01 &  0.69  &  43.83 &  0.90  &   4.96 &  0.05  &  0.09 & 0.02  \\
2020 KT54 & 77 & 7 &  26.51 &  0.24  &  40.18 & 0.07  &  36.56 &  0.77  &  40.74 &  0.78  &  18.30 &  0.18  &  0.10 & 0.03  \\
2020 KT55 & 29 & 6 &  24.98 &  0.36  &  43.56 & 0.29  &  31.26 &  6.14  &  62.30 & 25.04  &   5.55 &  0.21  &  0.46 & 0.28  \\
2020 KU11* & 162 & 6 &  26.57 &  0.41  &  50.71 & 0.04  &  36.40 &  0.54  &  44.31 &  0.48  &  15.48 &  0.07  &  0.18 & 0.02  \\
2020 KU53 & 77 & 5 &  25.25 &  0.19  &  47.84 & 0.11  &  39.87 &  1.78  &  44.19 &  0.64  &   2.93 &  0.01  &  0.10 & 0.03  \\
2020 KU54 & 75 & 5 &  26.39 &  0.15  &  38.37 & 0.10  &  38.17 &  0.21  &  45.02 &  0.97  &   6.67 &  0.08  &  0.15 & 0.02  \\
2020 KU55 & 76 & 6 &  25.83 &  0.20  &  44.39 & 0.09  &  43.05 &  0.97  &  44.65 &  0.81  &   5.49 &  0.06  &  0.04 & 0.02  \\
2020 KV11* & 794 & 9 &  25.29 &  0.31  &  65.94 & 0.00  &  35.96 &  0.00  &  95.87 &  0.04  &   4.64 &  0.00  &  0.62 & 0.00  \\
2020 KV53 & 80 & 7 &  25.56 &  0.53  &  44.87 & 0.07  &  40.94 &  1.87  &  45.15 &  0.36  &   4.37 &  0.03  &  0.09 & 0.04  \\
2020 KV54 & 26 & 5 &  26.11 &  0.35  &  43.37 & 0.30  &  29.95 &  6.55  &  81.14 & 58.00  &   4.80 &  0.15  &  0.52 & 0.29  \\
2020 KV55 & 78 & 9 &  24.72 &  0.19  &  41.78 & 0.04  &  38.36 &  0.38  &  41.58 &  0.41  &  17.43 &  0.10  &  0.08 & 0.02  \\
2020 KW11* & 160 & 6 &  25.75 &  0.30  &  48.71 & 0.00  &  39.93 &  0.15  &  45.20 &  0.04  &   4.80 &  0.00  &  0.12 & 0.00  \\
2020 KW53 & 77 & 7 &  25.96 &  0.21  &  46.52 & 0.10  &  41.38 &  0.95  &  44.35 &  0.77  &   5.15 &  0.06  &  0.07 & 0.02  \\
2020 KW54 & 75 & 5 &  26.38 &  0.20  &  43.29 & 0.14  &  42.85 &  0.52  &  182.55 & 33.88  &   6.20 &  0.08  &  0.76 & 0.05  \\
2020 KW55 & 24 & 4 &  25.10 &  0.21  &  40.33 & 0.33  &  28.74 &  5.86  &  65.87 & 32.94  &   2.39 &  0.04  &  0.50 & 0.29  \\
2020 KX11 & 77 & 5 &  26.09 &  0.56  &  74.81 & 0.29  &   4.11 &  0.35  &  110.07 & 42.70  &  55.52 &  4.45  &  0.96 & 0.01  \\
2020 KX53 & 30 & 5 &  25.99 &  0.55  &  46.54 & 0.30  &  29.69 &  5.93  &  66.21 & 31.36  &   4.09 &  0.11  &  0.50 & 0.28  \\
2020 KX54 & 75 & 6 &  26.44 &  0.13  &  44.09 & 0.10  &  41.49 &  0.99  &  43.92 &  0.97  &   2.89 &  0.02  &  0.05 & 0.03  \\
2020 KX55* & 422 & 11 &  24.21 &  0.21  &  35.55 & 0.00  &  35.47 &  0.00  &  67.01 &  0.00  &   2.73 &  0.00  &  0.47 & 0.00  \\
2020 KY11 & 3 & 2 &  25.35 &  0.76  &  55.34 & 3.91  &  13.28 & 12.90  &  67.72 & 47.70  &  25.37 & 18.44  &  0.84 & 0.14  \\
2020 KY53* & 427 & 9 &  25.07 &  0.28  &  30.06 & 0.00  &  23.26 &  0.01  &  48.77 &  0.03  &   4.72 &  0.00  &  0.52 & 0.00  \\
2020 KY54 & 27 & 4 &  26.52 &  0.22  &  43.17 & 0.38  &  30.87 &  6.53  &  75.91 & 42.57  &   4.12 &  0.13  &  0.51 & 0.28  \\
2020 KY55 & 79 & 7 &  25.72 &  0.35  &  39.20 & 0.06  &  38.60 &  0.44  &  46.41 &  0.92  &   3.46 &  0.02  &  0.17 & 0.03  \\
2020 KZ53 & 77 & 6 &  25.73 &  0.30  &  37.43 & 0.08  &  37.38 &  0.06  &  59.76 &  1.50  &   4.09 &  0.03  &  0.37 & 0.02  \\
2020 KZ54 & 27 & 5 &  26.16 &  0.18  &  48.60 & 0.42  &  28.30 &  6.30  &  80.48 & 57.97  &  10.56 &  0.58  &  0.54 & 0.28  \\
2020 KZ55 & 77 & 8 &  24.27 &  0.19  &  37.16 & 0.03  &  36.82 &  0.18  &  40.72 &  0.28  &  17.13 &  0.07  &  0.10 & 0.01  \\
2020 LA21 & 25 & 4 &  25.99 &  0.25  &  42.82 & 0.38  &  29.97 &  6.38  &  68.54 & 33.79  &   3.56 &  0.09  &  0.50 & 0.28  \\
2020 LB21 & 73 & 3 &  26.12 &  0.37  &  45.35 & 0.18  &  16.42 &  1.28  &  46.27 &  3.19  &   6.89 &  0.20  &  0.64 & 0.05  \\
2020 MA53 & 56 & 5 &  24.01 &  0.21  &  28.90 & 0.08  &  28.59 &  0.14  &  40.48 &  1.02  &   2.11 &  0.01  &  0.29 & 0.02  \\
2020 MB53 & 54 & 4 &  25.33 &  0.18  &  36.50 & 0.45  &  36.10 &  0.52  &  48.39 &  5.96  &  16.71 &  1.07  &  0.25 & 0.10  \\
2020 MC53 & 6 & 3 &  25.02 &  0.15  &  49.66 & 2.73  &  15.49 & 12.62  &  60.08 & 36.98  &   8.54 &  3.33  &  0.76 & 0.21  \\
2020 MD53 & 54 & 4 &  25.78 &  0.29  &  40.12 & 0.64  &  37.10 &  2.57  &  46.22 &  6.68  &   4.76 &  0.35  &  0.21 & 0.10  \\
2020 ME53 & 54 & 4 &  25.84 &  0.18  &  44.07 & 0.73  &  34.38 &  6.24  &  42.26 &  5.50  &   3.59 &  0.23  &  0.20 & 0.10  \\
2020 MF53 & 6 & 3 &  25.75 &  0.16  &  59.62 & 3.00  &  20.65 & 13.92  &  78.98 & 54.62  &  41.13 & 19.11  &  0.73 & 0.18  \\
2020 MG53 & 51 & 3 &  25.76 &  0.16  &  45.26 & 0.46  &  29.06 &  6.45  &  66.87 & 32.45  &   5.03 &  0.29  &  0.51 & 0.28  \\
2020 MH53 & 76 & 8 &  25.58 &  0.23  &  41.72 & 0.08  &  40.65 &  0.85  &  44.60 &  0.57  &   2.45 &  0.01  &  0.09 & 0.01  \\
2020 MJ53 & 4 & 2 &  26.25 &  0.44  &  87.72 & 4.06  &  23.91 & 24.35  &  114.66 & 89.28  &  14.91 &  9.30  &  0.84 & 0.15  \\
2020 MK53 & 4 & 2 &  25.99 &  0.49  &  162.03 & 5.35  &  46.54 & 45.03  &  212.81 & 161.42  &  31.44 & 20.77  &  0.83 & 0.15  \\
2020 ML53 & 4 & 2 &  26.40 &  0.21  &  48.89 & 3.04  &  11.82 & 12.76  &  54.78 & 33.74  &   7.52 &  4.57  &  0.86 & 0.15  \\
2020 MM53 & 54 & 4 &  26.47 &  0.18  &  42.77 & 1.08  &  26.48 &  8.77  &  38.26 &  5.41  &   2.11 &  0.15  &  0.33 & 0.16  \\
2020 MN53 & 4 & 2 &  26.44 &  0.37  &  41.94 & 2.59  &   9.67 &  6.51  &  49.16 & 30.42  &  41.54 & 20.82  &  0.79 & 0.15  \\
2020 MO53 & 4 & 2 &  26.27 &  0.19  &  54.61 & 3.29  &  14.53 & 15.04  &  74.21 & 61.77  &   3.26 &  0.81  &  0.85 & 0.16  \\
2020 MP53 & 52 & 3 &  26.40 &  0.48  &  53.23 & 2.16  &   9.73 &  7.01  &  39.47 &  9.31  &  25.56 & 11.55  &  0.74 & 0.19  \\
2020 MQ53 & 56 & 5 &  25.98 &  0.39  &  56.75 & 0.59  &  55.69 &  1.17  &  979.04 & 660.60  &  73.94 &  2.99  &  0.93 & 0.03  \\
2020 MY52 & 54 & 4 &  25.53 &  0.17  &  39.18 & 0.75  &  34.91 &  3.27  &  42.21 &  6.04  &  22.80 &  2.49  &  0.19 & 0.09  \\
2020 MZ52 & 52 & 3 &  26.00 &  0.26  &  48.67 & 1.90  &  23.58 & 12.11  &  40.68 &  7.85  &  16.57 &  4.66  &  0.47 & 0.22  \\
2020 PC95 & 3 & 2 &  24.94 &  2.19  &  40.53 & 3.16  &   5.64 &  5.42  &  43.99 & 27.60  &  21.20 & 16.35  &  0.87 & 0.12  \\
2020 PD95 & 3 & 2 &  25.39 &  1.72  &  80.08 & 4.74  &  22.11 & 20.41  &  98.18 & 66.82  &  35.50 & 24.97  &  0.83 & 0.14  \\
2020 PE95 & 3 & 2 &  25.68 &  0.99  &  46.09 & 3.60  &  11.48 & 11.69  &  53.48 & 34.57  &  14.53 &  8.80  &  0.85 & 0.15  \\
2020 PF95 & 3 & 2 &  25.83 &  0.36  &  54.65 & 3.98  &  15.25 & 12.58  &  66.88 & 44.20  &  41.92 & 25.64  &  0.81 & 0.14  \\
2020 PG95 & 3 & 2 &  26.55 &  0.18  &  49.24 & 3.70  &  10.30 & 11.27  &  53.47 & 31.85  &   7.89 &  5.66  &  0.87 & 0.14  \\
2021 LA44 & 501 & 7 &  23.34 &  0.43  &  38.40 & 0.00  &  31.58 &  0.13  &  39.44 &  0.08  &   4.79 &  0.00  &  0.20 & 0.00  \\
2021 LA45 & 89 & 3 &  24.11 &  0.28  &  43.22 & 0.21  &  40.02 &  1.78  &  42.39 &  1.54  &   1.64 &  0.01  &  0.06 & 0.03  \\
2021 LB44 & 880 & 9 &  25.14 &  0.50  &  43.58 & 0.00  &  37.68 &  0.04  &  42.24 &  0.01  &   2.72 &  0.00  &  0.11 & 0.00  \\
2021 LB45 & 499 & 5 &  25.45 &  0.56  &  36.60 & 0.00  &  36.38 &  0.11  &  47.75 &  0.11  &   9.04 &  0.00  &  0.24 & 0.00  \\
2021 LC44 & 3042 & 9 &  25.54 &  0.58  &  49.39 & 0.00  &  40.61 &  0.01  &  46.47 &  0.00  &   2.43 &  0.00  &  0.13 & 0.00  \\
2021 LD44 & 878 & 9 &  24.11 &  0.44  &  39.96 & 0.00  &  27.48 &  0.01  &  39.40 &  0.01  &   1.80 &  0.00  &  0.30 & 0.00  \\
2021 LE44 & 501 & 7 &  25.00 &  0.55  &  44.97 & 0.00  &  40.70 &  0.57  &  44.29 &  0.10  &   2.86 &  0.00  &  0.08 & 0.01  \\
2021 LF44 & 89 & 4 &  25.50 &  0.29  &  38.03 & 0.07  &  37.11 &  0.77  &  41.84 &  0.78  &  17.78 &  0.17  &  0.11 & 0.03  \\
2021 LG44 & 89 & 4 &  25.93 &  0.24  &  44.06 & 0.11  &  40.86 &  2.37  &  43.87 &  0.50  &   0.96 &  0.01  &  0.07 & 0.05  \\
2021 LH44 & 501 & 7 &  25.42 &  0.50  &  42.32 & 0.00  &  42.19 &  0.12  &  44.16 &  0.02  &   2.06 &  0.00  &  0.04 & 0.00  \\
2021 LJ44 & 501 & 6 &  25.01 &  0.54  &  44.50 & 0.00  &  41.80 &  0.58  &  44.25 &  0.06  &   3.99 &  0.00  &  0.06 & 0.01  \\
2021 LK44 & 89 & 4 &  24.54 &  0.25  &  43.41 & 0.05  &  37.01 &  0.89  &  87.31 &  3.43  &  16.17 &  0.10  &  0.58 & 0.03  \\
2021 LL44 & 89 & 4 &  25.45 &  0.32  &  42.33 & 0.08  &  40.92 &  1.11  &  42.86 &  0.61  &   1.80 &  0.01  &  0.05 & 0.03  \\
2021 LM44 & 501 & 7 &  24.95 &  0.61  &  47.56 & 0.00  &  40.08 &  0.18  &  43.96 &  0.03  &   1.35 &  0.00  &  0.09 & 0.00  \\
2021 LN44 & 501 & 5 &  24.51 &  0.41  &  32.69 & 0.00  &  31.38 &  0.08  &  39.33 &  0.05  &   5.19 &  0.00  &  0.20 & 0.00  \\
2021 LO44 & 468 & 7 &  24.94 &  0.28  &  35.27 & 0.00  &  29.47 &  0.11  &  39.26 &  0.09  &  18.57 &  0.00  &  0.25 & 0.00  \\
2021 LP44 & 468 & 7 &  24.48 &  0.33  &  35.43 & 0.00  &  31.79 &  0.10  &  55.42 &  0.29  &   4.54 &  0.00  &  0.43 & 0.00  \\
2021 LQ44 & 501 & 6 &  23.60 &  0.40  &  51.86 & 0.00  &  43.29 &  0.07  &  47.62 &  0.02  &  20.89 &  0.01  &  0.09 & 0.00  \\
2021 LR44 & 501 & 6 &  24.55 &  0.64  &  38.69 & 0.00  &  32.62 &  0.23  &  39.50 &  0.11  &   7.35 &  0.00  &  0.17 & 0.01  \\
2021 LS44 & 501 & 7 &  24.91 &  0.50  &  43.83 & 0.00  &  42.75 &  0.01  &  43.30 &  0.00  &   2.18 &  0.00  &  0.01 & 0.00  \\
2021 LT44 & 89 & 3 &  25.77 &  0.41  &  41.62 & 0.48  &  39.19 &  1.82  &  47.58 &  5.35  &   4.40 &  0.27  &  0.18 & 0.11  \\
2021 LU43 & 880 & 6 &  22.42 &  0.40  &  42.60 & 0.00  &  42.00 &  0.02  &  42.84 &  0.00  &   3.13 &  0.00  &  0.02 & 0.00  \\
2021 LU44 & 89 & 3 &  26.22 &  0.45  &  36.53 & 0.10  &   4.47 &  0.18  &  30.63 &  1.99  &  44.54 &  1.23  &  0.85 & 0.02  \\
2021 LV43 & 501 & 7 &  25.95 &  0.40  &  38.65 & 0.00  &  33.01 &  0.35  &  47.52 &  0.46  &   1.68 &  0.00  &  0.31 & 0.01  \\
2021 LV44 & 501 & 5 &  24.40 &  0.47  &  32.80 & 0.00  &  32.25 &  0.10  &  39.57 &  0.05  &   6.85 &  0.00  &  0.18 & 0.00  \\
2021 LW43 & 499 & 6 &  25.80 &  0.50  &  43.83 & 0.00  &  43.75 &  0.04  &  62.37 &  0.04  &  24.05 &  0.00  &  0.30 & 0.00  \\
2021 LW44 & 501 & 6 &  24.35 &  0.61  &  45.23 & 0.01  &  41.12 &  0.74  &  44.03 &  0.10  &   2.43 &  0.00  &  0.07 & 0.02  \\
2021 LX43 & 501 & 7 &  25.47 &  0.59  &  43.69 & 0.00  &  42.55 &  0.50  &  43.46 &  0.02  &   1.99 &  0.00  &  0.02 & 0.01  \\
2021 LX44 & 501 & 6 &  22.72 &  0.52  &  36.56 & 0.00  &  34.01 &  0.20  &  39.62 &  0.08  &   3.18 &  0.00  &  0.14 & 0.01  \\
2021 LY43 & 499 & 6 &  24.72 &  0.44  &  40.96 & 0.00  &  39.98 &  0.25  &  46.05 &  0.08  &   5.27 &  0.00  &  0.13 & 0.01  \\
2021 LY44 & 501 & 6 &  25.20 &  0.60  &  44.93 & 0.01  &  35.80 &  0.77  &  58.83 &  1.63  &   4.13 &  0.00  &  0.39 & 0.03  \\
2021 LZ43 & 501 & 7 &  23.65 &  0.49  &  38.67 & 0.00  &  38.56 &  0.04  &  43.60 &  0.02  &   1.49 &  0.00  &  0.12 & 0.00  \\
2021 LZ44 & 501 & 5 &  24.59 &  0.51  &  42.10 & 0.00  &  41.92 &  0.18  &  43.83 &  0.01  &   1.65 &  0.00  &  0.04 & 0.00  \\
2021 MA23 & 491 & 5 &  24.44 &  0.47  &  41.91 & 0.00  &  41.78 &  0.10  &  45.62 &  0.03  &   2.02 &  0.00  &  0.08 & 0.00  \\
2021 MW22 & 468 & 4 &  23.50 &  0.21  &  30.70 & 0.00  &  30.00 &  0.04  &  39.54 &  0.03  &   7.76 &  0.00  &  0.24 & 0.00  \\
2021 MX22 & 493 & 5 &  23.65 &  0.58  &  30.71 & 0.00  &  30.65 &  0.01  &  39.52 &  0.01  &   6.47 &  0.00  &  0.22 & 0.00  \\
2021 MY22 & 491 & 5 &  24.37 &  0.60  &  30.16 & 0.00  &  29.96 &  0.03  &  39.51 &  0.03  &   3.00 &  0.00  &  0.24 & 0.00  \\
2021 MZ22 & 80 & 3 &  24.79 &  0.30  &  41.27 & 0.19  &  38.82 &  2.26  &  45.20 &  0.82  &   3.67 &  0.10  &  0.15 & 0.06  \\
2021 NS71 & 63 & 2 &  24.98 &  0.32  &  48.91 & 2.33  &  13.44 &  8.72  &  51.90 & 26.21  &  10.05 &  3.77  &  0.73 & 0.20  \\
2022 LA14 & 29 & 4 &  26.29 &  0.34  &  46.62 & 0.83  &  28.53 &  7.54  &  72.77 & 40.39  &   1.60 &  0.19  &  0.54 & 0.27  \\
2022 LA15 & 465 & 6 &  25.63 &  0.47  &  39.19 & 0.01  &  38.70 &  0.36  &  42.00 &  0.09  &  19.64 &  0.01  &  0.08 & 0.01  \\
2022 LA16 & 28 & 3 &  25.66 &  0.31  &  39.55 & 0.58  &  29.20 &  5.93  &  83.00 & 58.56  &   1.24 &  0.03  &  0.54 & 0.27  \\
2022 LA17 & 464 & 5 &  25.53 &  0.60  &  41.80 & 0.04  &  39.62 &  2.01  &  46.22 &  0.59  &   2.16 &  0.00  &  0.14 & 0.05  \\
2022 LB14 & 465 & 6 &  24.17 &  0.51  &  43.21 & 0.01  &  39.57 &  1.72  &  43.05 &  0.32  &   2.26 &  0.00  &  0.08 & 0.05  \\
2022 LB15 & 465 & 6 &  24.95 &  0.50  &  32.01 & 0.01  &  31.87 &  0.12  &  39.80 &  0.14  &   5.48 &  0.00  &  0.20 & 0.01  \\
2022 LB16 & 29 & 4 &  25.47 &  0.17  &  32.90 & 0.42  &  25.45 &  4.75  &  62.21 & 31.69  &   6.21 &  0.41  &  0.53 & 0.26  \\
2022 LB17 & 2 & 2 &  26.20 &  0.22  &  46.55 & 3.60  &   9.87 & 10.83  &  51.69 & 32.30  &   7.86 &  4.97  &  0.87 & 0.14  \\
2022 LC14 & 465 & 6 &  25.53 &  0.56  &  37.36 & 0.02  &  33.92 &  2.11  &  45.01 &  1.83  &   1.76 &  0.00  &  0.25 & 0.08  \\
2022 LC15 & 465 & 5 &  26.19 &  0.53  &  47.63 & 0.08  &  36.53 &  4.89  &  49.09 &  3.66  &   1.77 &  0.01  &  0.26 & 0.17  \\
2022 LC16 & 29 & 4 &  24.84 &  0.25  &  32.71 & 0.37  &  25.74 &  4.29  &  72.49 & 52.08  &   8.12 &  0.48  &  0.54 & 0.26  \\
2022 LD14 & 298 & 5 &  24.54 &  0.20  &  44.79 & 0.04  &  38.32 &  4.32  &  47.63 &  2.26  &   2.29 &  0.01  &  0.20 & 0.14  \\
2022 LD15 & 29 & 3 &  26.04 &  0.21  &  43.69 & 0.72  &  29.49 &  6.84  &  68.02 & 32.28  &   2.76 &  0.21  &  0.52 & 0.27  \\
2022 LD16 & 464 & 4 &  25.30 &  0.54  &  41.96 & 0.04  &  37.39 &  3.74  &  47.69 &  2.24  &   2.27 &  0.01  &  0.22 & 0.12  \\
2022 LE14 & 28 & 3 &  26.20 &  0.13  &  46.19 & 0.84  &  31.06 &  7.44  &  91.70 & 72.60  &   1.10 &  0.13  &  0.54 & 0.27  \\
2022 LE15 & 29 & 3 &  26.16 &  0.17  &  35.20 & 0.49  &  29.78 &  3.84  &  101.60 & 72.16  &  20.58 &  1.35  &  0.62 & 0.21  \\
2022 LE16 & 465 & 6 &  24.48 &  0.41  &  32.99 & 0.00  &  30.40 &  0.30  &  39.46 &  0.26  &  10.87 &  0.00  &  0.23 & 0.01  \\
2022 LE17 & 464 & 5 &  25.37 &  0.21  &  43.72 & 0.06  &  33.13 &  5.94  &  53.19 & 12.48  &   1.24 &  0.01  &  0.37 & 0.26  \\
2022 LF14 & 465 & 5 &  25.21 &  0.78  &  37.15 & 0.01  &  36.86 &  0.25  &  44.82 &  0.08  &  13.10 &  0.01  &  0.18 & 0.01  \\
2022 LF15 & 29 & 3 &  26.55 &  0.44  &  50.10 & 1.20  &   4.91 &  2.90  &  52.89 & 40.57  &  13.14 &  4.05  &  0.89 & 0.09  \\
2022 LF16 & 29 & 4 &  23.96 &  0.16  &  51.22 & 0.47  &  29.46 &  5.97  &  71.60 & 36.63  &  26.63 &  1.77  &  0.53 & 0.26  \\
2022 LG14 & 465 & 5 &  24.61 &  0.46  &  35.34 & 0.00  &  23.65 &  0.03  &  12792.44 & 11980.15  &  11.66 &  0.00  &  1.00 & 0.00  \\
2022 LG15 & 465 & 6 &  25.15 &  0.63  &  43.01 & 0.03  &  40.41 &  2.14  &  45.05 &  0.45  &   1.53 &  0.00  &  0.10 & 0.06  \\
2022 LG16 & 29 & 4 &  25.37 &  0.20  &  40.15 & 0.51  &  30.96 &  5.70  &  74.42 & 36.74  &  28.48 &  1.96  &  0.53 & 0.26  \\
2022 LH14 & 29 & 3 &  25.71 &  0.17  &  44.13 & 0.68  &  29.42 &  6.81  &  72.37 & 40.78  &   5.93 &  0.55  &  0.52 & 0.27  \\
2022 LH15 & 465 & 6 &  22.77 &  0.58  &  43.55 & 0.01  &  42.66 &  0.81  &  45.59 &  0.12  &   1.24 &  0.00  &  0.06 & 0.02  \\
2022 LH16 & 29 & 4 &  23.55 &  0.22  &  41.53 & 0.34  &  30.16 &  6.41  &  86.36 & 65.14  &   0.48 &  0.00  &  0.53 & 0.28  \\
2022 LJ14 & 465 & 5 &  26.23 &  0.49  &  40.30 & 0.05  &  35.12 &  3.51  &  45.20 &  2.43  &   1.97 &  0.01  &  0.23 & 0.13  \\
2022 LJ15 & 465 & 6 &  24.80 &  0.56  &  44.87 & 0.02  &  41.65 &  2.14  &  45.14 &  0.31  &   3.26 &  0.00  &  0.08 & 0.06  \\
2022 LJ16 & 29 & 4 &  25.21 &  0.22  &  31.82 & 0.39  &  24.11 &  4.60  &  54.83 & 25.81  &  12.01 &  0.80  &  0.51 & 0.27  \\
2022 LK14 & 465 & 5 &  24.10 &  0.62  &  48.95 & 0.01  &  47.06 &  0.26  &  148.87 &  3.23  &  32.57 &  0.01  &  0.68 & 0.01  \\
2022 LK15 & 465 & 5 &  24.43 &  0.53  &  42.22 & 0.02  &  40.94 &  1.19  &  46.34 &  0.25  &   3.05 &  0.00  &  0.12 & 0.03  \\
2022 LK16 & 29 & 4 &  22.28 &  0.15  &  37.10 & 0.30  &  30.90 &  4.54  &  66.31 & 25.68  &  31.16 &  1.33  &  0.50 & 0.23  \\
2022 LL14 & 465 & 6 &  23.51 &  0.51  &  48.68 & 0.01  &  37.32 &  0.37  &  43.33 &  0.11  &   9.26 &  0.01  &  0.14 & 0.01  \\
2022 LL16 & 464 & 5 &  25.20 &  0.51  &  34.54 & 0.03  &  29.93 &  3.78  &  55.55 & 12.46  &   0.63 &  0.00  &  0.45 & 0.19  \\
2022 LM14 & 735 & 7 &  24.39 &  0.55  &  46.98 & 0.00  &  45.12 &  0.12  &  46.76 &  0.01  &   2.55 &  0.00  &  0.04 & 0.00  \\
2022 LM15 & 465 & 6 &  24.83 &  0.63  &  42.44 & 0.02  &  40.34 &  1.73  &  45.98 &  0.40  &   3.77 &  0.00  &  0.12 & 0.05  \\
2022 LM16 & 465 & 6 &  25.97 &  0.56  &  50.60 & 0.06  &  37.43 &  2.61  &  46.61 &  1.20  &   3.58 &  0.01  &  0.20 & 0.08  \\
2022 LN14 & 465 & 5 &  24.61 &  0.41  &  41.06 & 0.02  &  38.30 &  2.44  &  45.23 &  0.76  &   1.89 &  0.00  &  0.15 & 0.07  \\
2022 LN15 & 29 & 4 &  24.56 &  0.16  &  34.84 & 0.37  &  25.12 &  5.01  &  53.70 & 23.55  &   5.48 &  0.34  &  0.49 & 0.28  \\
2022 LN16 & 465 & 5 &  25.30 &  0.58  &  33.74 & 0.00  &  33.40 &  0.31  &  39.54 &  0.19  &   6.87 &  0.01  &  0.16 & 0.01  \\
2022 LO14 & 464 & 4 &  23.67 &  0.43  &  40.81 & 0.05  &  36.10 &  3.80  &  46.09 &  2.27  &   0.76 &  0.00  &  0.22 & 0.12  \\
2022 LO15 & 29 & 4 &  24.89 &  0.20  &  36.86 & 0.39  &  30.00 &  4.64  &  101.84 & 95.40  &   5.61 &  0.32  &  0.58 & 0.24  \\
2022 LO16 & 465 & 6 &  25.60 &  0.60  &  35.43 & 0.01  &  33.71 &  0.85  &  64.77 &  3.22  &   3.96 &  0.01  &  0.48 & 0.04  \\
2022 LO17 & 735 & 7 &  24.53 &  0.63  &  46.25 & 0.00  &  41.90 &  0.00  &  44.08 &  0.00  &   2.34 &  0.00  &  0.05 & 0.00  \\
2022 LP14 & 29 & 4 &  26.12 &  0.33  &  42.13 & 0.64  &  26.91 &  6.28  &  72.71 & 50.22  &   0.68 &  0.02  &  0.53 & 0.28  \\
2022 LP15 & 465 & 6 &  24.88 &  0.44  &  32.52 & 0.01  &  32.36 &  0.17  &  36.44 &  0.03  &   4.22 &  0.00  &  0.11 & 0.01  \\
2022 LP16 & 28 & 3 &  23.32 &  0.15  &  42.09 & 0.35  &  29.58 &  6.22  &  68.49 & 35.00  &   2.75 &  0.12  &  0.50 & 0.28  \\
2022 LP17 & 28 & 3 &  25.87 &  0.28  &  46.01 & 1.18  &  26.11 &  8.49  &  77.29 & 54.65  &   2.06 &  0.18  &  0.57 & 0.26  \\
2022 LQ14 & 465 & 5 &  24.81 &  0.51  &  38.35 & 0.00  &  34.34 &  0.78  &  58.27 &  1.68  &   6.95 &  0.00  &  0.41 & 0.03  \\
2022 LQ15 & 29 & 4 &  26.48 &  0.33  &  42.89 & 0.67  &  30.99 &  6.71  &  92.88 & 74.45  &   2.70 &  0.23  &  0.54 & 0.27  \\
2022 LQ16 & 29 & 4 &  26.43 &  0.48  &  76.81 & 2.04  &  44.22 & 16.60  &  147.52 & 123.79  &   7.89 &  1.47  &  0.60 & 0.25  \\
2022 LR14 & 465 & 5 &  25.24 &  0.60  &  45.74 & 0.03  &  40.06 &  1.85  &  44.30 &  0.35  &   4.97 &  0.01  &  0.10 & 0.05  \\
2022 LR15 & 28 & 2 &  26.32 &  0.35  &  74.28 & 3.90  &  21.73 & 18.89  &  86.64 & 52.83  &  41.00 & 24.45  &  0.80 & 0.15  \\
2022 LR16 & 465 & 6 &  25.60 &  0.53  &  43.34 & 0.02  &  38.84 &  2.20  &  42.95 &  0.42  &   4.65 &  0.01  &  0.10 & 0.06  \\
2022 LS14 & 465 & 5 &  24.85 &  0.36  &  48.23 & 0.05  &  38.45 &  3.36  &  46.86 &  1.33  &   0.82 &  0.00  &  0.18 & 0.10  \\
2022 LS15 & 29 & 4 &  26.36 &  0.29  &  47.37 & 0.79  &  23.65 &  6.27  &  64.99 & 37.35  &   0.70 &  0.00  &  0.57 & 0.24  \\
2022 LS16 & 29 & 2 &  25.59 &  0.94  &  87.65 & 4.20  &  27.74 & 21.95  &  122.65 & 97.52  &  44.19 & 23.82  &  0.79 & 0.16  \\
2022 LT14 & 28 & 2 &  26.66 &  0.19  &  64.82 & 2.88  &  36.33 & 10.22  &  109.47 & 77.79  &  58.01 & 15.72  &  0.59 & 0.24  \\
2022 LT15 & 735 & 5 &  23.80 &  0.41  &  42.51 & 0.00  &  42.31 &  0.05  &  43.41 &  0.00  &   5.26 &  0.00  &  0.03 & 0.00  \\
2022 LT16 & 28 & 3 &  25.79 &  1.29  &  72.81 & 2.64  &  29.30 & 17.25  &  105.16 & 76.29  &   7.04 &  2.01  &  0.68 & 0.24  \\
2022 LU14 & 28 & 3 &  24.86 &  0.17  &  35.40 & 0.44  &  29.36 &  4.29  &  82.34 & 42.67  &   0.73 &  0.00  &  0.59 & 0.22  \\
2022 LU15 & 28 & 3 &  25.96 &  0.40  &  38.33 & 0.58  &  30.05 &  5.36  &  92.75 & 68.80  &   6.19 &  0.51  &  0.57 & 0.25  \\
2022 LU16 & 28 & 2 &  26.12 &  0.52  &  71.48 & 3.65  &  19.78 & 18.61  &  84.48 & 54.48  &  28.29 & 19.05  &  0.82 & 0.16  \\
2022 LV13 & 29 & 4 &  24.95 &  0.32  &  42.01 & 0.48  &  28.20 &  6.24  &  69.11 & 40.46  &   1.16 &  0.08  &  0.51 & 0.28  \\
2022 LV14 & 28 & 3 &  25.50 &  0.24  &  45.64 & 0.71  &  29.31 &  6.92  &  78.83 & 52.67  &   0.89 &  0.05  &  0.54 & 0.28  \\
2022 LV15 & 465 & 6 &  25.85 &  0.58  &  44.45 & 0.04  &  38.17 &  3.33  &  44.87 &  1.13  &   3.03 &  0.01  &  0.15 & 0.10  \\
2022 LV16 & 465 & 5 &  24.51 &  0.54  &  44.88 & 0.05  &  34.32 &  5.66  &  50.96 &  8.13  &   0.97 &  0.01  &  0.34 & 0.24  \\
2022 LW13 & 465 & 6 &  24.80 &  0.44  &  31.95 & 0.00  &  31.92 &  0.03  &  39.43 &  0.01  &  11.54 &  0.00  &  0.19 & 0.00  \\
2022 LW14 & 465 & 5 &  25.64 &  0.71  &  49.57 & 0.07  &  36.06 &  3.34  &  46.62 &  1.89  &   1.09 &  0.00  &  0.23 & 0.11  \\
2022 LW16 & 465 & 5 &  25.84 &  0.63  &  42.17 & 0.05  &  36.06 &  3.94  &  45.02 &  1.85  &   1.52 &  0.00  &  0.20 & 0.13  \\
2022 LX13 & 29 & 4 &  24.32 &  0.19  &  45.08 & 0.45  &  29.80 &  6.34  &  89.05 & 80.13  &   1.95 &  0.12  &  0.52 & 0.29  \\
2022 LX14 & 29 & 4 &  25.11 &  0.27  &  44.77 & 0.51  &  28.99 &  6.52  &  66.82 & 32.06  &   1.53 &  0.10  &  0.51 & 0.28  \\
2022 LX15 & 28 & 3 &  25.97 &  0.52  &  82.15 & 1.56  &   9.13 &  2.09  &  83.13 & 53.49  &  58.07 & 16.12  &  0.86 & 0.08  \\
2022 LX16 & 464 & 5 &  23.79 &  0.62  &  46.83 & 0.02  &  43.58 &  1.94  &  47.17 &  0.32  &   4.48 &  0.01  &  0.08 & 0.05  \\
2022 LY13 & 465 & 6 &  25.21 &  0.63  &  34.96 & 0.00  &  33.93 &  0.36  &  39.63 &  0.17  &  14.41 &  0.01  &  0.14 & 0.01  \\
2022 LY14 & 465 & 5 &  26.17 &  0.66  &  39.89 & 0.04  &  37.54 &  2.16  &  45.87 &  0.86  &   1.88 &  0.00  &  0.18 & 0.06  \\
2022 LY15 & 27 & 2 &  25.64 &  1.00  &  76.37 & 3.80  &  23.13 & 19.94  &  116.38 & 108.28  &  31.94 & 19.89  &  0.80 & 0.17  \\
2022 LY16 & 464 & 5 &  26.05 &  0.46  &  46.27 & 0.07  &  38.26 &  3.86  &  46.14 &  1.42  &   1.93 &  0.01  &  0.17 & 0.12  \\
2022 LZ13 & 29 & 4 &  25.72 &  0.27  &  45.15 & 0.65  &  27.77 &  6.62  &  65.13 & 31.70  &   1.07 &  0.08  &  0.52 & 0.27  \\
2022 LZ14 & 29 & 4 &  25.32 &  0.25  &  30.09 & 0.38  &  24.44 &  3.84  &  72.57 & 55.73  &  11.28 &  0.70  &  0.56 & 0.24  \\
2022 LZ15 & 465 & 5 &  24.71 &  0.54  &  40.32 & 0.02  &  38.46 &  1.42  &  48.05 &  0.70  &   3.53 &  0.01  &  0.20 & 0.04  \\
2022 LZ16 & 463 & 4 &  25.05 &  0.55  &  41.46 & 0.02  &  39.05 &  2.14  &  44.64 &  0.52  &   3.39 &  0.01  &  0.13 & 0.06  \\ \enddata
\tablecomments{
List of TNOs discovered in our 2020, 2021, and 2022 Subaru-HSC searches.
These discoveries have been submitted to the Minor Planet Center. 
See Section~\ref{sec:linking} for details on the orbit fitting. Arc and Epochs denote the span of each arc, and number of different nights on which the object was detected and linked. Mag and $\sigma_{mag}$ are an estimate of the source brightness and its uncertainty. $a$, inc, and e are the fitted orbit semi-major axis, inclination, and eccentricity, and $\sigma_a$, $\sigma_{inc}$ and $\sigma_{e}$ uncertainties in each, respectively. Objects marked with a * were successfully recovered with follow-up observations with HST. Attempted recoveries of 2020 KY11 and 2020 KB56 with HST were unsuccessful. 
}
\end{deluxetable*}

\section{Colour Transforms for the EB-gri filter \label{sec:col_trans}}

\begin{figure}
    \includegraphics[width=0.95\textwidth]{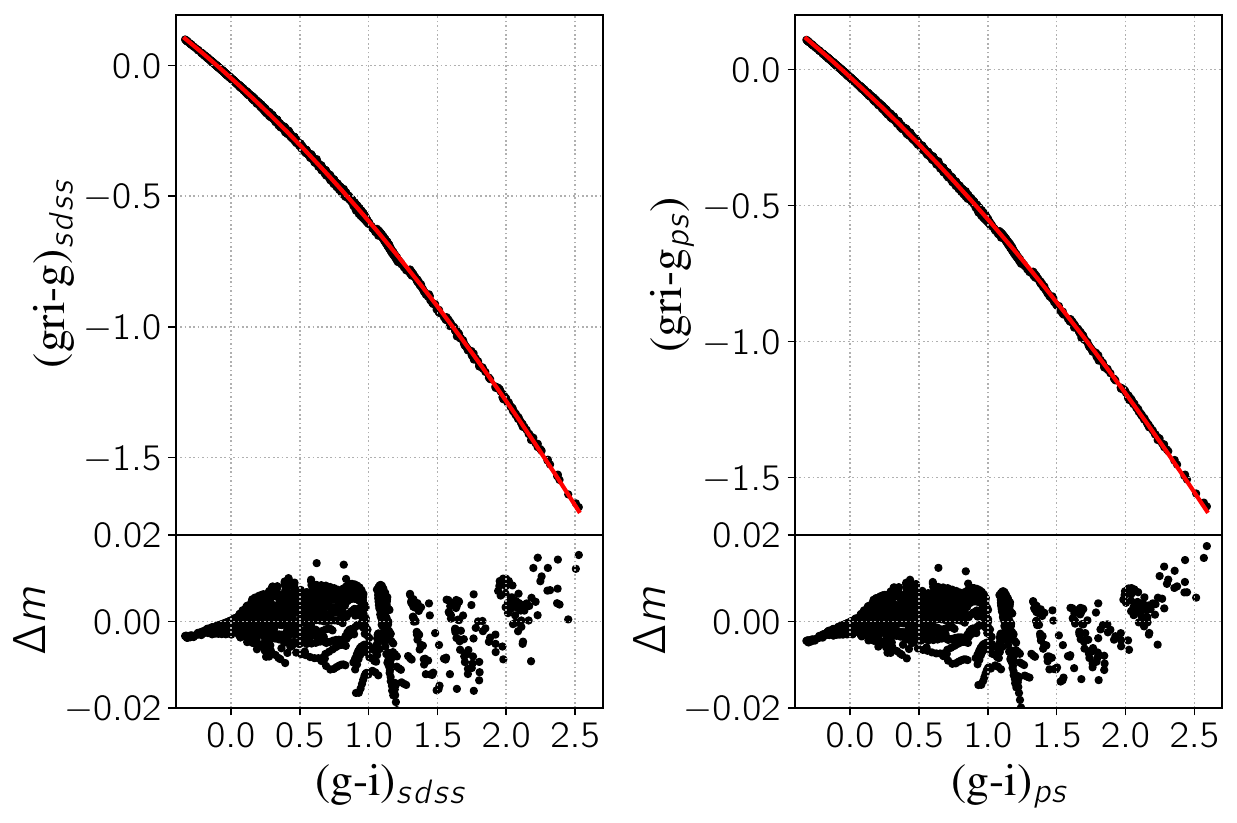}
    \caption{Estimated colour terms for the HSC EB-gri filter for the SDSS and PS systems. The bottom panel shows the difference between the modelled colors and the 3rd degree polynomial fit to the colors shown by the red curve. \label{fig:col-trans}}
\end{figure}

In Figure~\ref{fig:col-trans} we present estimates of the color transformations between the (g-i) and the wide filter (gri-g) for the SDSS and PS filter systems (for clarity we forgo including ``EB'' in equations). These estimates were made with with \emph{pysynphot} package. For each model stellar spectrum included in the \emph{pysynphot} data files, the (g-i) and (gri-g) colors were estimated using the transmission provided by the manufacturer of the EB-gri filter (see Figure~\ref{fig:filters}) and the appropriate filter transmission curves for the other filters. 3rd degree polynomials were fit to the outputs, resulting in the color terms:

\begin{align}
(gri-g)_\textrm{sdss} &= -0.047 -0.475(g-i)_\textrm{sdss} -0.0713(g-i)_\textrm{sdss}^2\\
(gri-g)_\textrm{ps} &= -0.027 -0.464(g-i)_\textrm{ps} -0.0587(g-i)_\textrm{ps}^2.
\end{align}
\noindent
For reference, the solar color $(g-i)_{\textrm{sdss}}=0.55$ translates to $(gri-g)_{\textrm{sdss}}=0.36.$

The above estimate does not include the transmission curve of the rest of the Subaru-HSC system,\footnote{\url{https://www.naoj.org/Projects/HSC/data/HSCWFCTx.dat}} which shows a $\sim8\%$ low-to-high variation across the EB-gri bandpass. This variation will somewhat affect the above result; we adopt a generous 0.05~mag uncertainty in the estimated EB-gri magnitude when converting tabulated (g-i) colors for photometric calibration purposes. 

\section{Simulated Orbit Fitting \label{sec:simorbitfit}}

\begin{figure}
    \includegraphics[width=0.95\textwidth]{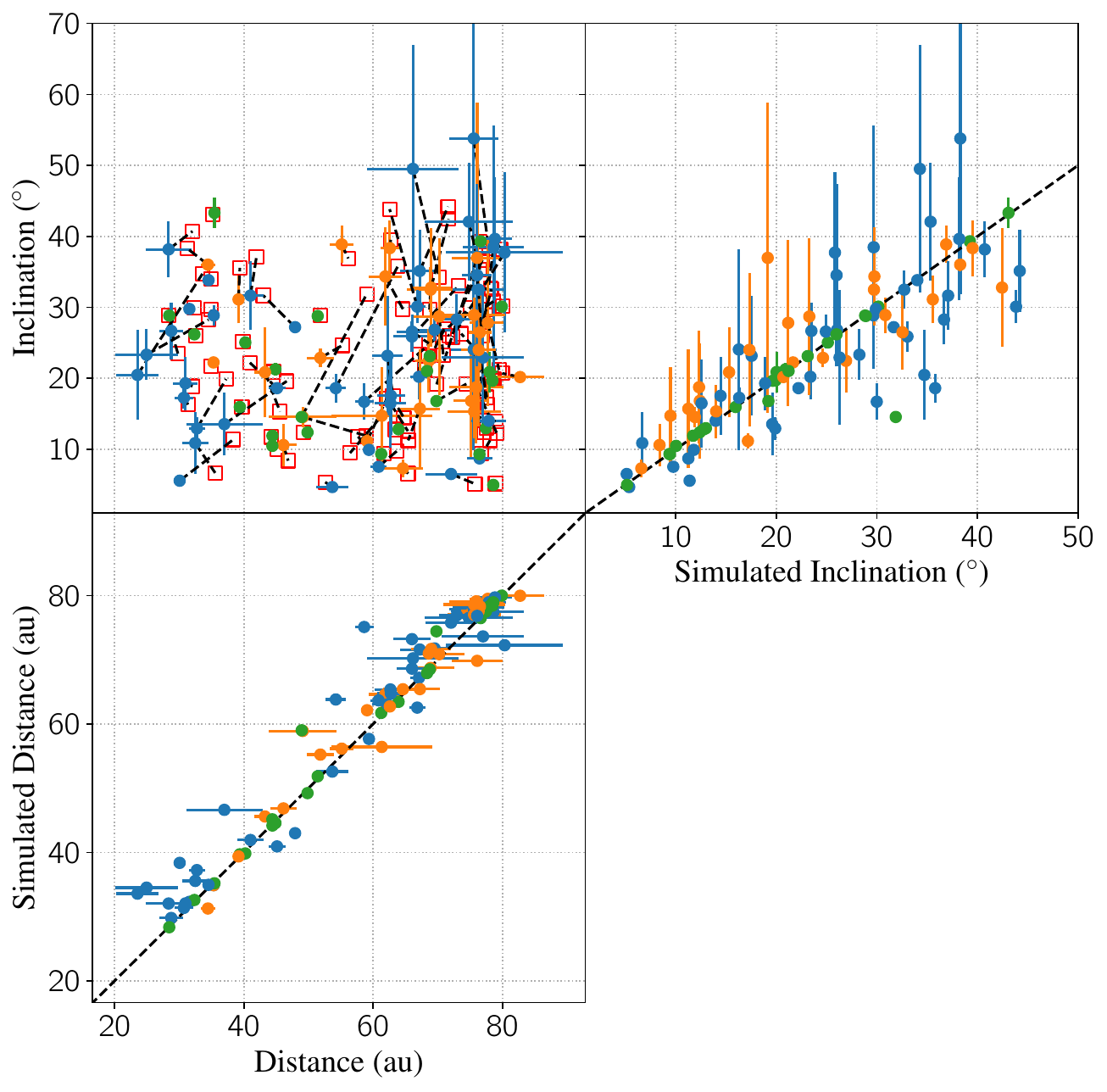}
    \caption{Results of orbit fitting to simulated ephemerides of objects in the 5:2 resonance model that fell within the \nh\ discovery field. \textbf{Top left:} Barycentric inclination vs. heliocentric distance. Fitted data are shown in circles with true values in red open squares. Simulated arcs of lengths of 2, 4, and 26 days are shown in blue, orange, and green respectively. The fitted and true values of each object are connected by black dashed lines. \textbf{Top right:} Fitted vs simulated barycentric inclinations for the objects shown top left. The 1-1 line is shown by the dashed line. \textbf{Bottom left:} Simulated vs. fitted heliocentric distances for the objects shown top left. The 1-1 line is shown by the dashed line. \label{fig:simorbitfits}}
\end{figure}

To assess the accuracy of our orbit-fitting scheme, we performed a set of orbit fits to simulated objects from the 5:2 resonance model discussed in Section~\ref{sec:discussion}. To do so, we selected those objects that fell within the survey footprint (see Table~\ref{tab:deteff}), and for each object simulated ephemerides for each date of observation. This large amount of synthetic data was then subsampled to mimic the detections we report and, in particular, the cases with short data arcs (see Table~\ref{tab:detections}). After a random synthetic object was selected, its ephemerides were reduced to a random subset that had less than four nights of data spanning no more than 40 days. There were sets of nights that fulfilled these conditions in 2021, 2022, and 2023, although not in 2020. The synthetic positions for the object were then turned into astrometry files for input into the orbit fitter, with randomized astrometry with an assumed uniform 150 milliarcsecond uncertainty (typical for low SNR detections in this data). The short data arcs were then fit, initially with a rough circular orbit to provide an initial condition, and then with the full MCMC estimator (see Section~\ref{sec:linking}) until it reached convergence. The results of the recovered orbits from fitting the synthetic data are shown in Figure~\ref{fig:simorbitfits}. The varied qualities of the simulated fitted orbits exhibit the same behaviour of our real detections, including the numerous short-arc links with matching large uncertainties. Importantly, the orbit fits to short-arc ephemerides tend to be somewhat biased to higher inclinations, in that the fitted values are higher than the true values beyond what the uncertainties reported by the fits would allow.

\end{document}